\documentclass[12pt, pa4paper]{article}
\usepackage[margin=1in]{geometry}

\usepackage[utf8x]{inputenc}
\usepackage[T1]{fontenc}
\usepackage{spverbatim}

\usepackage{graphicx}
\usepackage[colorlinks=true, allcolors=black]{hyperref}
\usepackage{float}
\usepackage{natbib}
\setcitestyle{authoryear,open={(},close={)}}
\usepackage{authblk}
\usepackage[T1]{fontenc}
\usepackage[T2A]{fontenc}
\usepackage{setspace}
\usepackage{amsmath}
\usepackage{breqn} 
\usepackage{float} 
\usepackage[compact]{titlesec}
\titlespacing{\section}{0pt}{2ex}{1ex}
\titlespacing{\subsection}{0pt}{1ex}{0ex}
\titlespacing{\subsubsection}{0pt}{0.5ex}{0ex}
\usepackage{url}
\usepackage{booktabs}
\usepackage{xcolor} 
\usepackage{color,soul} 
\usepackage{cleveref}
\usepackage{comment}
\usepackage{tabularx}

\newcommand{\pmi}{\mathrm{PMI} }

\begin{document}

\begin{center}
\thispagestyle{empty}
\parskip=18pt%
\vspace*{4\parskip}%

{\Large Invisible Women in Digital Diplomacy: A Multidimensional Framework for Online Gender Bias Against Women Ambassadors Worldwide \par}
\end{center}

\noindent Yevgeniy Golovchenko, Karolina Sta\'nczak, Rebecca Adler-Nissen, Patrice Wangen, \newline Isabelle Augenstein
\newline
\newline
\newline

\normalsize
\begin{abstract}	
Despite mounting evidence that women in foreign policy often bear the brunt of online hostility, the extent of online gender bias against diplomats remains unexplored. This paper offers the first global analysis of the treatment of women diplomats on social media. Introducing a multidimensional and multilingual methodology for studying online gender bias, it focuses on three critical elements: gendered language, negativity in tweets directed at diplomats, and the visibility of women diplomats. Our unique dataset encompasses ambassadors from 164 countries, their tweets, and the direct responses to these tweets in 65 different languages. Using automated content and sentiment analysis, our findings reveal a crucial gender bias. The language in responses to diplomatic tweets is only mildly gendered and largely pertains to international affairs and, generally, women ambassadors do not receive more negative reactions to their tweets than men, yet the pronounced discrepancy in online visibility stands out as a significant form of gender bias. Women receive a staggering 66.4\% fewer retweets than men. By unraveling the invisibility that obscures women diplomats on social media, we hope to spark further research on online bias in international politics.


\end{abstract}

\doublespacing
\section{Introduction}
Foreign policy has long been a domain reserved for men. When former US Ambassador to the UN, Samantha Power, was first appointed to President Barack Obama's administration to work on the National Security Council, she hid the fact that she was pregnant because she believed that it would be an impediment to her prospects. Then when she took the job, she was offered lower pay and a smaller office than her men counterparts \citep{barrington_women_2020}. While women such as Hillary Clinton, Condoleezza Rice, and Madeleine Albright have served as foreign ministers, most current ambassadors are identifying as men \citep{towns_gender_2017}. Historically, it is only relatively recently that women were allowed entrance to the diplomatic corps. While the world is becoming more gender-inclusive, diplomacy remains rife with gender inequalities and discriminatory practices, making it difficult for women to enter diplomacy at the highest position. Women in diplomacy are discriminated against in a variety of ways, having to hold themselves to higher standards than men \citep{neumann_body_2008}, having to conduct themselves differently and think more about their appearances than their men counterparts \citep{towns_diplomacy_2020}, or -- as was the case in many countries until the 1970s -- to remain unmarried if they were to keep a diplomatic post (see \citet{mccarthy_women_2014}). Overall, women are seen as less capable of being on the front line and dealing with national security issues, damaging the prestige and foreign policy of a country. Such perceptions may have consequences not just for gender equality in foreign policy, but also for the policy being conducted. Yet we still know little about how gender inequalities translate across countries and on the key social media platform for foreign policy: Twitter (for a recent exception, see \citet{jezierska_incredibly_2021}). We know that on social media, influential women face significant online hate, from dismissive insults to gendered sexual harassment \citep{kumar_mapping_2021}. This paper therefore asks: What is the character and scope of gender bias on social media targeted toward women ambassadors? 

We focus specifically on Twitter, recently rebranded as X. As it was called Twitter at the time of data collection and analysis, we will continue to refer to the platform as Twitter. This focus is justified by Twitter's status as the predominant social media platform used by diplomats globally, preferred over other platforms such as Facebook, Snapchat, and Instagram in this particular domain \citep{adler2022blended}. Twitter allows foreign ministers and diplomats to promote their views and policies by engaging with the audiences directly. Moreover, established news media around the world frequently amplify public officials on Twitter, including diplomats, by quoting their tweets in news articles. 

We explore gender bias by examining whether women ambassadors (1) are targeted with more negativity (2) are approached with gendered language (i.e., grounded in gender stereotypes) and (3) are less visible online, in tweets compared to their men colleagues. 
We construct a dataset consisting of the Twitter accounts of ambassadors from 164 UN member states and the several million tweets written in 65 languages directed directly at them. Using automated content analysis, Natural Language Processing (NLP), including its subsets, sentiment analysis, and gendered language detection, we investigate the scope and nature of digital gender biases against women and analyze the factors that help explain these biases and inequality. We use this unique global dataset to develop a multidimensional and multilingual approach to gender bias. 

Our central finding is that online bias against women ambassadors is not primarily rooted in outright negativity, such as uncivil comments or negative tones. Contrary to widespread belief, women do not face a heightened degree of negativity in the public responses to their tweets on a global scale. Moreover, while there exists a minor gendered aspect in the language used in direct Twitter replies to women ambassadors, it is not of substantial magnitude. Importantly, most public responses to women ambassadors revolve around matters related to foreign affairs and diplomacy, rather than resorting to discussions of physical appearance or perpetuating gendered stereotypes. Instead, the primary source of online bias against women ambassadors stems from a distinct lack of online visibility. This manifests itself in women ambassadors receiving significantly fewer retweets than their men colleagues. This gender bias is more subtle in nature, unfolding through unseen mechanisms rather than overtly visible content. However, this bias is of paramount importance, as visibility is a fundamental prerequisite for engaging in public diplomacy and ranks as one of the most vital resources on social media platforms. Our identification of this subtle bias carries significant implications for addressing online bias. On one hand, the diplomatic Twittersphere may present a `safer' online space for women compared to other political domains, and on the other hand, it underscores the deeply ingrained nature of these biases in our language, including tweets by women ambassadors. These biases may prove more challenging to confront and ultimately overcome.


The paper proceeds as follows. First, in \Cref{sec:framework}, we discuss the gaps in the existing literature on diplomacy, gender, and online harassment and develop a multidimensional conceptualization of gender bias focusing on three critical aspects that we term online visibility, gendered language, and negativity. Second, we develop our specific hypotheses about how gender bias plays out in digital diplomacy worldwide in \Cref{sec:expectations}. Third, we describe our research design, data collection, and methods in Sections \ref{sec:design}, \ref{sec:data}, and \ref{sec:method}. Fourth, \Cref{sec:results} presents our analysis and findings on the types of biases against women diplomats and how they relate to nationality, country prestige, and women diplomats' own tweeting behavior. Finally, in \Cref{sec:conclusion}, we discuss the implications of our research, its scope conditions, and circumstances under which gender bias might further impact the landscape of online diplomacy and international politics.

\section{Online Gender Bias and Diplomacy: A Multidimensional Framework}
\label{sec:framework}

Although social media have become one of the key platforms for foreign policy communication and diplomacy, little is known about stereotypes or biases that social media users communicate about women diplomats. Studying gender bias against women diplomats on social media is important as diplomacy plays a crucial role in shaping international relations and policy. Gender bias may undermine the participation and influence of women in diplomatic efforts, limiting their contributions to global issues. Moreover, online gender bias directed at women diplomats can harm their professional reputation, credibility, and effectiveness. In this section, we will highlight how our study seeks to fill the gaps and contribute to our existing knowledge by developing a multidimensional and multilingual approach to gender bias on social media. 

To the best of our knowledge, there are no scholarly studies specifically addressing gender bias against diplomats on social media. We know from existing largely qualitative work that women have experienced exclusion and a range of biases when they work as diplomats \citep{sluga2015women,rahman2017women,mccarthy2014women,erlandsson2019,davey2019}. More recently, 
several large-scale studies have demonstrated profound inequalities and discrimination in diplomacy. \citet{towns2016gender} collected the first comprehensive dataset of 7,000 ambassador appointments from the fifty highest GDP-ranked countries of 2014 and it shows that women are still less likely to occupy high-status positions compared to their counterparts. They find that despite the recent emergence of women into the field of diplomacy in large numbers, 85\% of ambassador postings are occupied by men, indicating an extremely tilted gender composition of the global diplomatic corps. Further solidifying this evidence of gender bias, \citet{towns_diplomacy_2020} show that women ambassadors from the US, UK, Denmark, and Sweden are less likely than men to be posted with states with higher economic status and countries with inter-state conflict and that these gender differences in ambassador appointments persist over time. Most recently, \citet{niklasson2023diplomatic} demonstrated that states generally tend to post more women ambassadors to countries that project gender equality in an attempt to signal value alignment and climb the international status hierarchy. 

If women diplomats also face gender bias on social media, it can further discourage aspiring women from entering the field. Other studies show that gender bias can sometimes escalate into online harassment, cyberbullying, or even threats \citep{griezel2012uncovering}.  But beyond women's representation, inclusion, and safety in diplomacy online, gender bias on social media can also affect international politics. It has been shown that the adoption of an explicit Feminist foreign policy shapes diplomatic discourse and practice \citep{aggestam_bergman-rosamond_2016,frohlich2023feminist}. Moreover,  diplomatic efforts often involve building and maintaining relationships with other nations, and within international organizations. Here, surveys have demonstrated that gender stereotypes impact negotiation styles among national diplomats in the EU \citep{naurin2019gender}. Gender bias can damage diplomatic relationships if it, for example,  leads to misunderstandings, tensions, or perceptions of disrespect. Yet, as \citet{aggestam2019gender} emphasize, we need more research to shift attention from North America and Europe to the entire world and we need to study gender bias online.

In the rest of this section, we begin to fill these gaps by examining gender bias online and at a global scale by drawing on broader literature on gender bias, social media, and politics to develop a multifaceted approach to gender bias. Below we distinguish between three aspects of online gender bias: visibility  (i.e., no retweets, low followership), gendered language (e.g., ``soft'', ``emotional''), and negativity (e.g., ``women are worthless''). The benefit of a multidimensional approach is that it captures distinct aspects of bias on social media. Sometimes, these biases are at work simultaneously, often reinforcing each other. Other times only one or two of these forms of discrimination can be observed. 


\subsection{Visibility 
}
While most people would list physical violence as the most extreme form of discrimination, being overlooked can also have severe consequences. One form of online gender bias lies in being ignored or disregarded. \citet{nilizadeh_twitters_2016} have examined over 94,000 Twitter users, and show the association between perceived gender and online visibility (understood as how often Twitter users are followed, assigned to lists, and retweeted). Women are less frequently followed and their posts are shared less often. 
In general, online texts about women are found to be consistently shorter and less often edited than those about men \citep{field2020controlled, nguyen2020}. On Wikipedia, articles about women are more likely to include links to articles about men than the other way round \citep{wagner2015its}. 
Moreover, users perceived as women experience a `glass ceiling', similar to the barrier women face in attaining higher positions in society, thus men tend to be among the top-followed users. Other studies, predominantly based on US Twitter data, confirm this observation. One study points to the most significant difference existing in the top 1\% of those most followed, where the difference is 15\% and then the difference decreases until the top 14\%, where the fraction of women becomes higher than the fraction of men \citep{messias_white_2017}. 
In the case of diplomacy, the ability to be heard or seen is crucial because visibility is one of the main power resources on social media as well as a prerequisite for carrying out digital diplomacy in the first place. We know from small n-surveys of Irish women diplomats that they felt excluded from exclusively male social networks and, as a result, particularly abroad, they tended to create their own support networks \citep{barrington_women_2020}. 

\subsection{Gendered Language
}
In this paper, we use the term `gendered language' to describe language usage with a bias towards a particular social gender, following \citet{bigler2015genered}. This would include using gender-specific terms referring to professions or people, such as `businessman' or `waitress', or using the masculine pronouns (he, him, his) to refer to people in general, such as `A doctor should know how to communicate with his patients'. The use of gendered language, like the examples above, perpetuates what \citet{jule2017beginner} calls ``the historical patriarchal hierarchy that has existed between men and women, where one (man) is considered the norm, and the other (woman) is marked as other – as something quite different from the norm''. Stereotypes around women and feminine speech tend to be stronger than those pertaining to men and masculinity, arguably because male speech is taken as  ``neutral'' or normative  (i.e., ``real speech'' \citep{quina_language_1987}). 
\citet{wagner2015its} first uncovered gendered language in Wikipedia biographies, revealing a higher likelihood of words corresponding to gender, relationships, and families being found in female Wikipedia pages rather than male ones. Further studies focusing on Wikipedia also identified a greater amount of content related to sex and marriage in female biographies \cite{graells2015}.
We know from other studies that gender traits biases put women at a disadvantage to men, when they candidates for political positions, since the qualities viewed most favorably by voters are those stereotypically associated with masculinity, including competence \citep{ksiazkiewicz_implicit_2018}, assertiveness, and self-confidence \citep{huddy_gender_1993}. Stereotypical feminine traits, such as compassion, warmth, and sensitivity, may be less valued, or only viewed as favorable on certain issues, such as healthcare or education \citep{ksiazkiewicz_implicit_2018}. 
While \citet{marjanovic2022gender} found equal public interest towards men and women politicians on Reddit, as measured by comment distribution and length, this interest may not be equally professional and reverent; female politicians are much more likely to be referenced using their first name and described in relation to their body, clothing, and family than male politicians. 

Within diplomacy, there are few studies of gendered language in the diplomatic profession, but one in-depth study drawing on interviews with queer women diplomats from Australia, shows that they struggle with a need to suppress their identity, and the personal challenges that came with navigating a particularly men-dominated and heteronormative field, resulted in self-censoring and opting out of many diplomatic appointments – the emotional and psychological toll falling heavily on women and queer individuals \citep{aggestam2019gender}.   

\subsection{Negativity}
In diplomacy, women's participation is challenged when they are exposed to verbal or physical assault. There is currently no available data on the number of women diplomats who have been assaulted, neither online nor offline. Nevertheless, news reports frequently reveal that diplomats are targeted with negative remarks or even physical attacks. A former American envoy faced harassment from a senior lawmaker while she was serving on the White House Security Council \citep{voa2017women}. In Australia, Japan, and most other countries, women have historically not been posted to hardship posts and dangerous regions because it would not be "appropriate" or "safe" given their gender, thus restricting their careers and status within the foreign service \citep{stephenson2019domestic,flowers2018women}. As long as certain diplomatic posts remain reserved for men and women diplomats are labeled as weak and untrustworthy, women will continue to be marginalized in diplomacy \citep{MinarovaBanjac2018GenderCI}. 

Whether such negative perceptions of women also translate into the online sphere is unknown. \cite{henry2020technology} found in their systematic review that women and gender-diverse individuals are more likely to experience online harassment, such as stalking, doxing, and non-consensual sharing of explicit content. They also underline that harassment on social media can have severe psychological and emotional consequences, leading to self-censorship and withdrawal from online spaces, ultimately limiting free expression.

However, negativity in language can also take more subtle forms than assault and microaggression, often evident in the overall tone of a text. \citet{mertens2019} observed systematic gender differences in the tone of tweets aimed at politicians. The study revealed that tweets directed at right-leaning women and left-leaning men were typically more positive, indicating nuanced variations in how language tone aligns with gender and political orientation.

\section{Expectations About Online Gender Bias Against Women Ambassadors}
\label{sec:expectations}
With this theoretical motivation to study the multidimensional character of gender bias in mind, this section specifies gender biases that we are likely to see and develops hypotheses concerning online gender bias against women ambassadors. These different forms of biases are all consequential. Our assumptions about online gender bias against women ambassadors resonate with various strands of scholarship (see \Cref{sec:framework}), which document the many direct and indirect ways through which women in politics are being made invisible both online and offline. 

\textbf{H1}: \textit{Women diplomats are less visible on Twitter than men diplomats.}

Building on research presented in \citet{haakansson_women_2021}, we expect that gender bias is mediated by media visibility. In line with this, we expect the following outcome:

\textbf{H1.1}: \textit{Gender bias expressed through visibility is stronger among diplomats who are assigned to countries with high prestige.}

\textbf{H2}: \textit{Women diplomats face more negative responses than their men counterparts}.

\textbf{H2.1} \textit{Gender bias expressed through negative tweets is stronger among diplomats with higher visibility on Twitter.}

Just as we would expect the response to women diplomats to be biased, we would also expect the women themselves to -- at least to some extent -- conform to the gendered inequality structures that they are embedded in. It is important to stress that women may also behave differently online than men in terms of the language they use. In general, research suggests that women tend to use more positive language than men, especially positive emotions \citep{kucuktunc_large-scale_2012, kivran-swaine_joy_2012, iosub_emotions_2014}. In contrast, men have been found to refer more to anger in their language use than women \citep{mehl_sounds_2003}. 
We expect that the qualities that the broader public perceives as important and beneficial when assessing diplomats align neatly with long-standing gender stereotypes. Twitter users may perceive men speakers and masculine speech patterns as higher in competence, but lower in social warmth. 

Studies of social media users have also shown that women use warmer,  more polite, and more deferential language, while men language use is more hostile, more impersonal, and more assertive \citep{cunha_he_2014, park_women_2016}. However, aware of these biases and of the tendency of women to reproduce these gendered stereotypes in their own tweeting,  women diplomats may also adapt their communicative strategies, emphasizing personality traits associated with masculinity while downplaying those considered feminine \citep{brooks_he_2013}. They may also adopt `counter stereotypic' behavior,  such as attacking opponents or eschewing emotional language. However, in doing so, they may risk backlash for being viewed as unlikeable or unladylike \citep{bauer_effects_2017, windett_gendered_2014}. Trapped in a double bind, women diplomats who attempt to counteract gender stereotypes may be perceived as neither ``leader nor... lady'', and punished for their failure to perform their gender in ways that conform to social expectations \citep[p.~279]{bauer_effects_2017}.

In line with this, we hypothesize gender bias to be mediated by the diplomats' own tweeting behavior \citep{nilizadeh2016twitter}. We expect that the gender bias against women is strongest when they do not conform to the stereotypical norms:  

\textbf{H2.2} \textit{Gender bias expressed through negative tweets increases when women write more negative tweets.}

\textbf{H3}: \textit{Diplomats are targeted with gendered language tweets.}

\section{Research Design}
\label{sec:design}

We employ a computational social science approach to investigate 981,562 multilingual retweets of ambassadors and 458,932 Twitter replies in 65 languages to 1,960 ambassadors on Twitter from 164 UN member states. Our study is not limited to any political topic but covers the entire online conversation of all diplomats when they use their official Twitter handles and thus are perceived as ambassadors.

We test all of the hypotheses by using regression models to examine whether the gender of the ambassadors, the main independent variable, correlates with \emph{visibility}, as well as the \emph{negativity} and the level of \emph{gendered language} of the content targeted towards them in replies on Twitter. We use state-of-the-art methods from Natural Language Processing (NLP) to measure both negativity and gendered language. These methods as well as the operationalization steps will be described in detail below. First, we will turn to the data collection process.


\section{Data}
\label{sec:data}
The dataset was collected through a three-step process. 
First, we collected the original data for this study from a list of ambassadors. 
In some instances, the data includes \textit{chargés d'affaires} who serve as head of mission in the temporary absence of the ambassador. Here, we identified ambassadors by consulting the leading reference source on international organizations and ambassadors: \textit{Europa World}'s digital archive of all UN member states registered in the world. The book version of the archive has been used in prior research on diplomacy such as \citet{bezerra_going_2015}, \citet{kinne2014dependent}, \citet{rhamey2013diplomatic} and \citet{volgy_major_2011} (please see \citet{niklasson2023diplomatic} for a comparison of their own data with Europa World Yearbooks). Using this invaluable resource, we initially identified ambassadorial postings for the vast majority of countries. In the few instances where ambassadorial postings were absent from the \textit{Europa World} archive during the data collection period (for Montenegro, UK, Serbia, US, and Canada), we conducted a manual search to ensure coverage. 

 \begin{figure}[h!]
 \caption{Number of ambassadors on Twitter by country of origin}
\centering
\includegraphics[width=0.95\textwidth]{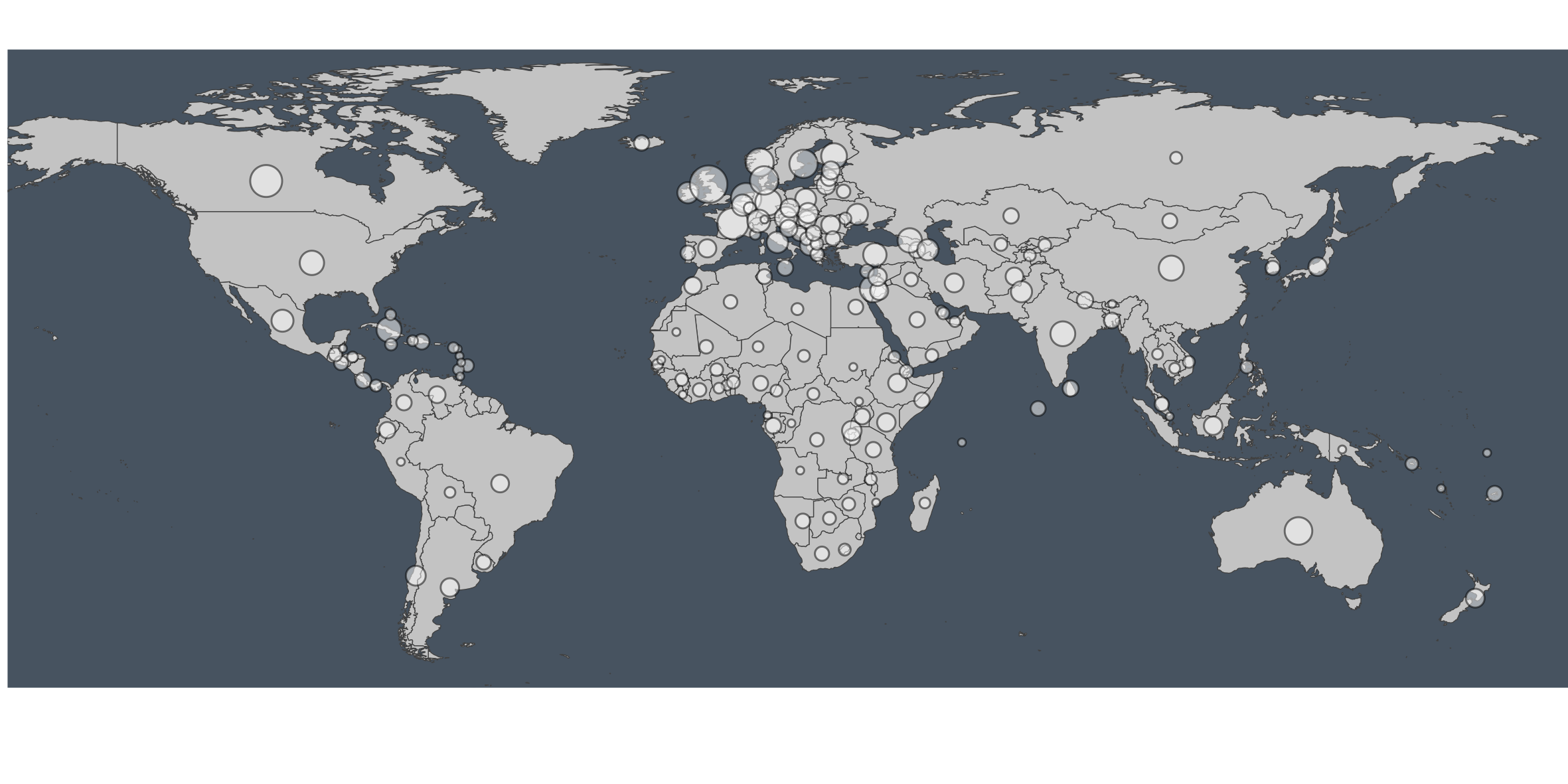}
\label{fig:amb_map}
\end{figure}


In the second phase, we conducted a comprehensive search for ambassadors on Twitter. To achieve this, six student assistants were provided with a detailed annotation guide, instructing them to (1) identify the public Twitter handles of the ambassadors and (2) deduce the publicly displayed gender through an examination of profile descriptions, recent posts, and profile images.
Initially, we employed an automated script to filter out names that did not correspond to any existing Twitter profiles. Subsequently, the remaining names were manually verified by the annotators by adjusting search parameters, such as removing middle names, if the initial full name search yielded no results (see Appendix for the annotation guide).

To secure the reliability of the annotation process, Cohen's kappa \citep{cohen1960kappa} coefficients were calculated, revealing substantial agreement among annotators. In the first stage, the agreement reached a Cohen's kappa of 0.94, while in the second stage, it rose to 0.98. These coefficients were derived from the analysis of 100 tweets annotated by all annotators in each respective stage.
It is important to note that our annotation guide transcends a binary men/women categorization. Annotators were instructed to identify any gender that the diplomats might use to describe themselves on their Twitter accounts. Despite this inclusive approach, after meticulous scrutiny, no ambassadors were found who identified as gender-non-binary. Consequently, our analysis is confined to the categories of women and men for pragmatic reasons.

In the third step, we collect tweets that are both posted by and directed at ambassadors. We use Twitter's REST API to download the ambassadors' tweets from their own timelines. In addition, we use Twitter's Search API to extract tweets that contain the diplomats' handle names to capture interactions with the ambassador accounts. This includes replies, mentions, as well as retweets. The tweets posted by the ambassadors were collected through the Twitter API from January 31 to May 17, 2021. 
Lastly, we use the Twitter Academic API\footnote{https://developer.twitter.com/en/use-cases/do-research/academic-research} to download historical tweets that contain the foreign minister/ambassador handle names as well as their own tweets for the countries that have been updated in June 2021: US, UK, Canada Serbia, and Montenegro. 

The resulting dataset consists of 1,960 ambassadors on Twitter from 164 UN member states. \Cref{fig:amb_map} shows the geographic distribution of these ambassadors by country of origin (i.e., the sending country). In total, our dataset consists of 458,932 replies in 65 languages to diplomatic actors as well as retweets of the same accounts from January 31 2021 to June 26 of the same year. To the best of our knowledge, this makes for the most complete list of individual ambassador accounts in academic research. 

In compliance with Twitter's data access policies, our dataset is limited to publicly available tweets. The results of negativity and gendered language cannot be generalized to private ``Direct Messages'', where one would expect to find more uncivil content. Nevertheless, public tweets are important, because they are much more visible than private messages and therefore have the potential to shape the public view of the ambassadors' and their work. Moreover, we do not distinguish between bias coming from human users and automated accounts or ``bots'' (see \citet{orabi2020detection} for an overview of the challenges with bot detection). From the point of view of the users, however, negative replies, gendered language, or the lack of visibility replies may be a real barrier, regardless of whether the issue originates from inauthentic accounts. 

On a user level, our data is limited to information about ambassadors and not their audiences. It is possible to infer the gender of thousands of ordinary users through automated tools to test, for instance, whether gender bias towards women ambassadors is mainly driven by men, who engage with the diplomats. We opted out of this option due to ethical concerns. By categorizing gender for a multinational set of users through automated (often gender-binary) tools, one may run the risk of putting accounts into man/woman categories (by design) with no non-binary option. By manually and carefully evaluating how ambassadors portray themselves online, we limit the risk of a priori excluding non-binary gender identities.

\section{Methodology}
\label{sec:method}
In the sections below, we explain the methodology employed in this study. We introduce the proxies used to measure gender bias: visibility, negativity, and gendered language (\Cref{sec:method-proxy}). 
We then describe the variables that are taken into consideration for control purposes in our analysis: diplomats' tweeting behavior and the prestige of the country they are sent to or received by (\Cref{sec:method-control}). 


\subsection{Proxies for Gender Bias}
\label{sec:method-proxy}
We define \emph{visibility}, \emph{negativity}, and \emph{gendered language} as our key variables of interest. These serve as proxies for gender bias, and in statistical terminology, are referred to as the dependent variables.

\subsubsection{Visibility} Visibility is measured in terms of the total number of retweets that a diplomatic actor has received during the data collection time period. 
We consider retweets as the main measure of visibility because retweets reflect active engagement with as well as active dissemination of ambassadors' tweets. 
As a supplementary measure, we operationalize visibility as the number of followers for the respective diplomatic accounts. 
However, user visibility through retweets is a more relevant metric for two reasons. Firstly, a high number of followers does not guarantee that the followers see or engage with the posted content. In contrast, retweets unequivocally signify a direct and active engagement with the content, further amplifying the original tweet's visibility each time it occurs. Secondly, ambassadors often inherit their Twitter accounts, including their followers, from their predecessors, who are predominantly men. Consequently, a woman ambassador's account may boast a substantial following, but this figure might predominantly reflect the accumulated visibility achieved by her men counterparts in the past. In sum, when assessing visibility exclusively through the lens of follower count, there is a perilous risk of overlooking the nuanced gender bias that women ambassadors may encounter. 

\subsubsection{Negativity} 

\begin{figure}[H]
 \caption{Sentiment classification examples}
\centering
\includegraphics[width=0.9\textwidth]{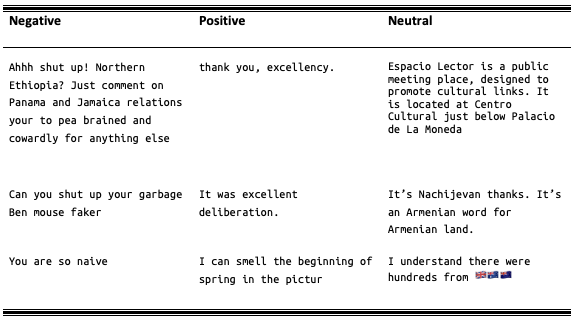}
\label{fig:send_examples}
\end{figure}

Negativity is quantified using the sentiment classifications from a multilingual XLM-RoBERTa-based language model \citep{DBLP:conf/nips/ConneauL19}, which was trained on roughly 200 mio. tweets and fine-tuned for multilingual sentiment analysis task in eight languages (Arabic, English, French, German, Hindi, Italian, Spanish, and Portuguese) \citep{barbieri2021xlmtwitter}.
We employ this model to classify each tweet in the curated dataset into one of the three sentiment categories: positive, neutral, and negative. We validate this model by comparing its results to the valence scores from the VAD lexicon \citep{mohammad-2018-obtaining} obtained for each tweet in a correlation analysis. In \Cref{fig:send_examples}, we include some examples of replies in English that have been classified either as positive, negative, or neutral in sentiment.

\subsubsection{Gendered language} Gendered language is operationalized in two ways. First, we use the NRC VAD Lexicon \citep{mohammad-2018-obtaining} to test whether online audiences use more dominant language in their replies to women ambassadors. Second, we use point-wise mutual information ($\pmi$) to examine to what extent words associated with replies to men and women follow a gender-stereotypical pattern.

We note that sentiment is indicative solely of hostile biases rather than more nuanced biases extant in language. Therefore, alongside sentiment, we analyze common words that are directed towards diplomats which can reveal more subtle biases such as benevolent sexism.  

\paragraph{Dominant Language}

First, we investigate the gendered language of the tweets operationalized as the dominance ratings to calculate the perceived power levels of tweets in response to diplomats. To this end, we employ the NRC VAD Lexicon \citep{mohammad-2018-obtaining}, which to our knowledge is by far the largest and most reliable multilingual affect lexicon spanning 100 languages, and has been widely applied in linguistic studies \citep{li2020content,Mendelsohn2020AFF}. 
The dominance score of each in-corpus word in the text is summated and averaged over the tweets' number of in-corpus words to determine the tweet's average dominance. 
Thus, we are able to compute dominance scores solely for tweets that include at least one word from the lexicon. 

\paragraph{Words Co-occurrences}

Additionally, we provide a general overview of the word and topic choice in tweets directed towards and by ambassadors in our dataset. 
First, we use point-wise mutual information ($\pmi$) as a measure of association between a word being used in response to an ambassador and an ambassador's gender.
In general, $\pmi$ is a measure of association that examines co-occurrences of two random variables and quantifies the amount of information we can learn about a specific variable from another. We treat generated words as bags of words and analyze the $\pmi$ between gender $g \in \mathcal{G} = \{\mathit{man}, \mathit{woman}\}$, as in the case of our dataset, and a word $w$ as:

\begin{equation}
    \pmi(g,w)=\log \frac{p(g,w)}{p(g)p(w)}
\end{equation}

In particular, $\pmi$ quantifies the difference between the co-occurrence probability of a word and gender compared to their joint probability if they were independent. If a word is more often associated with gender, its $\pmi$ will be positive; if less, it will be negative.
For instance, we would expect a high $\pmi$ value for the pair $\pmi(\mathit{woman}, \mathit{pregnant})$ because their co-occurrence probability is greater than the independent probabilities of $\mathit{woman}$ and $\mathit{pregnant}$.
Therefore, in an ideally unbiased context, words like $\mathit{successful}$ or $\mathit{intelligent}$ would be expected to have a $\pmi$ value of approximately zero for all genders.


\subsection{Diplomat's Tweeting Behavior and Country's Prestige}
\label{sec:method-control}

In this study, we are interested in the effect of an ambassador's gender on the three discussed gender bias dimensions: visibility, negativity of replies, and dominant language of replies. In order to correctly estimate the effect of gender and avoid omitting relevant variables, we additionally include a diplomat's own tweeting behavior, their (receiving or sending) country, and this country's prestige.     

In selected regression models, we control for the country that sends the ambassador and the receiving country that the ambassador is assigned to. We refer to the two types as ``sending'' and ``receiving'' host country, respectively. 
Furthermore, we control for individual ambassador-level variables such as the \emph{activity}, measured as the logged total number of original tweets posted by the ambassadors during the data collection period. 
Lastly, we use a network approach to examine the \emph{prestige} of the ambassadors' position. We operationalize the latter by measuring the standardized in-degree (ranging from 0 to 1) of their host country in a network of diplomatic ties (see \citet{kinne2014dependent} for a similar operationalization and \citet{manor2019towards} for an overview of network prestige in digital diplomacy)). The higher the proportion of all the countries in the diplomatic network with an established embassy in the respective host country, the higher its prestige score. The network is constructed using the online version of the \textit{Europa World Year Book} and includes diplomatic missions where the ambassadors are not present on Twitter. 

We observe that there are instances where ambassadors are assigned to multiple host countries. 
The 1,960 ambassadors in the dataset are assigned to 2,389 diplomatic postings in total -- equivalent to 1.22 postings per ambassador. Approximately 15.38\% of all the women ambassadors on Twitter are assigned to more than one country, whereas the number is 9.74\% for men. This difference is both substantively large and statistically significant ($p<0.001$). These findings align with those presented by \citet{towns_gender_2017} in their study of (offline) ambassador appointments. \citet{towns_gender_2017} argue that women are more likely to be sent to multiple smaller embassies countries, which in itself could be interpreted as an indication that women are appointed to less prestigious positions.

\section{Results}
\label{sec:results}
\subsection{Visibility}

 \begin{figure}[H]
\centering
\caption{Retweets, negative replies, and dominance received by ambassadors. All four figures show descriptive means with 95\% confidence intervals and without controls. Figure \textbf{C} shows the mean dominance score \emph{per tweet}, while the remaining figures show the mean number of retweets and proportion of negative replies \emph{per ambassador}.}
\includegraphics[width=0.9\textwidth]{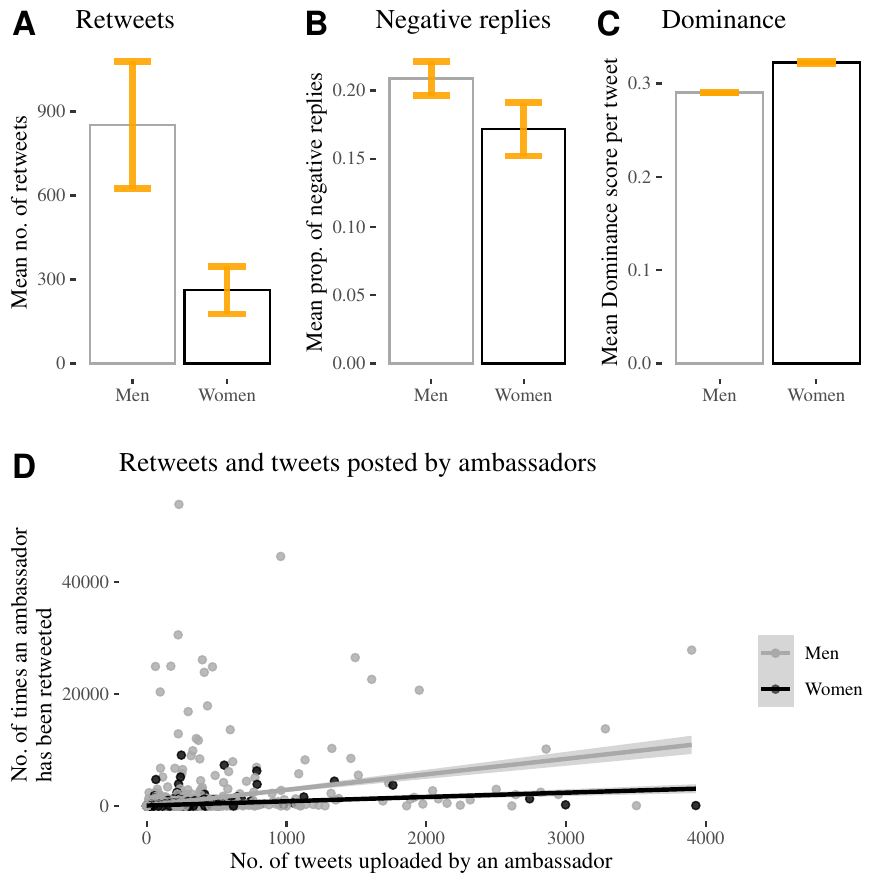}

\label{fig:mbassador_neg_dom_ret}
\end{figure}

We begin the analysis by testing Hypothesis 1, which predicts that women ambassadors are less visible online compared to their men colleagues. We find that this is indeed the case when measuring visibility as the number of retweets, as illustrated in the descriptive \Cref{fig:mbassador_neg_dom_ret} A. Women receive on average 406 fewer retweets ($p<0.05$) than men when excluding control variables. In other words, women receive on average 66.4\% fewer retweets. The difference in visibility through retweets is persistent even when examining ambassadors who themselves are highly active on Twitter. This is illustrated in \Cref{fig:mbassador_neg_dom_ret} D. Here, each node represents an ambassador, the axes reflect the number of times an ambassador is retweeted and the number of original tweets uploaded by the ambassadors themselves.

It is important to note that 19.5\% of women and 27.8\% of all men in the sample received 0 retweets during the time period. While the exact reason for this is unknown, the lack of retweets likely reflects inactivity on Twitter. Ambassadors who have not been retweeted a single time post 16.4 times fewer original tweets (13.7 tweets on average) than those who have posted at least one tweet. Our estimations of gender bias are therefore conservative when considering that there are more inactive ambassador men with 0 retweets than women. In other words, men would have an even higher average than women if one removed all inactive accounts from the data.

The difference in visibility between men and women becomes smaller when including control variables, however, it remains substantial. \Cref{table:reg_vis_multi} in the Appendix shows the results based on a negative binomial regression model
, in order to control for the sending countries (where the ambassadors are from) and receiving countries (where the ambassadors are assigned to). Using a multilevel framework, all of the ambassadors (i.e., observations) are nested either in their receiving country (Model 1, Model 3) or sending country (Model 2 and 4) with the countries serving as random effects in the models. Results in Model 1 in \Cref{table:reg_vis_multi} indicate that women ambassadors receive on average 45.7\% fewer retweets than their men colleagues.\footnote{The percentage is derived from the coefficient estimate: $(exp(-0.61)-1)*100=45.7$}This is the case when controlling for their receiving country, the number of tweets uploaded by the ambassador (\emph{n tweets}) and the global prestige of the receiving country, measured as the standardized number of countries that send an embassy to the respective country (\emph{in-degree (receiving country)}). 

The gender difference is smaller when controlling for sending instead of the receiving countries (Model 2). Here, women receive 36.2\% fewer retweets than men. In both cases, the difference is both statistically significant and substantively large. The findings are robust also when using a zero-inflated reiteration of the models as well as negative binomial models without the multilevel, nested data structure.

Estimates in Model 3 and 4 in \Cref{table:reg_vis_multi} indicate that women ambassadors have 16.5\% or 17.5\% fewer followers than men when controlling for the receiving country and sending country respectively. These results, however, are not robust. The difference is no longer statistically significant when using country-level fixed effects in the negative binomial models instead of nested, multilevel structure (see Appendix \Cref{table:reg_vis_multi_fixed}).

\paragraph{Visibility and Prestige}
We now turn to Hypothesis 1.1, which predicts that gender bias against women in terms of diminished visibility is more pronounced among ambassadors holding more prestigious positions. As previously mentioned, we analyze prestige by using standardized in-degree centrality, which measures the extent to which other nations establish embassies in the respective country. To examine this hypothesis, we employ a multilevel negative binomial regression model, where we control for the log number of tweets uploaded by the ambassador, the ambassador's receiving country (Model 1, Model 3), and the sending country (Model 2 and 4). The regression tables are available in Appendix \Cref{table:reg_vis_mediated_in-degree}.  

Our estimates for differences in retweets support the hypothesis: Women sent to prestigious countries (with above-median in-degree score) receive on average 42.9\% fewer retweets than women assigned to less prestigious destinations with up to median in-degree scores when controlling for the sending country.\footnote{The percentages are derived from the coefficient estimate: $(exp(-0.56)-1)*100=42.9$} The difference is at 36.2\% when controlling for the sending countries instead. In line with the results in the previous section, we observe no statistically significant difference when looking at the number of followers (Models 3 and 4 in Appendix \Cref{table:reg_vis_mediated_in-degree}).

In other words, we observe an online glass ceiling effect when examining visibility through retweets. The relatively few women who gain prestigious, diplomatic positions in the men-dominated diplomatic sphere may experience an additional barrier at the top in the competition for online visibility.  

\subsection{Negativity in Replies}

We now proceed to test Hypothesis 2, which predicts that women receive more negativity in public replies than men overall. We find no support for the hypothesis on a global level. This part of the analysis is limited to 1,424 ambassadors who have received at least one reply in the data. The descriptive \Cref{fig:mbassador_neg_dom_ret} B illustrates the proportion of negative replies sent to men and women. Contrary to the hypothesis, women receive on average 0.37 \emph{fewer} percent points of negative replies, when running a simple OLS without controls, as shown in Model 1 in Appendix \Cref{table:simple_OLS}. Although statistically significant ($p<0.01$), the difference is arguably too small to be substantively meaningful. 

 \begin{figure}[t]
\centering
\caption{The difference in the proportion of negative and positive replies sent to the ambassadors. Figure \textbf{A} and Figure \textbf{B} show estimates with receiving and sending country fixed effects respectively. The brackets mark the relevant hypotheses corresponding to each row. For estimates with receiving country fixed effects (Figure \textbf{A}), the first row (H2) is based on Model 1, the second row (H2.1) is based on Model 2, and the third row is based on Model 3 (H2.2). For estimates with sending country fixed effects (Figure \textbf{B}), the estimates are based on models 4, 5, and 6 in the respective order. All of the models are available in Appendix \Cref{table:neg_sentiment_fixed} and \Cref{table:pos_sentiment_fixed}.}
\includegraphics[width=0.99\textwidth]{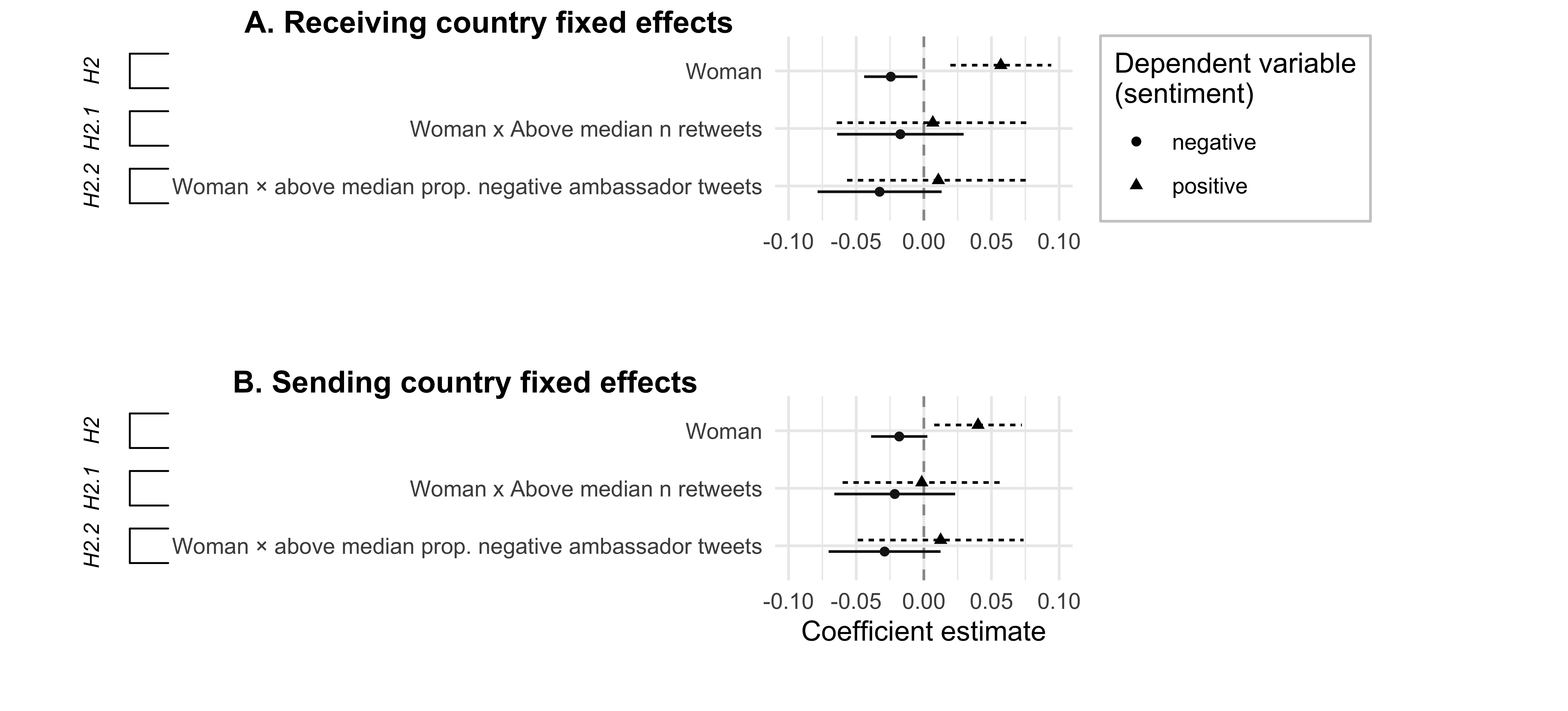}

\label{fig:dot_wiskers_sentiment}
\end{figure}

The first row in \Cref{fig:dot_wiskers_sentiment} shows the difference in proportion of negative replies between men and women when controlling for \emph{number of tweets} uploaded by the ambassadors, the \emph{in-degree} of the receiving country, a dummy variable for whether the ambassadors receive \emph{above median number of retweets} as well as countries through receiving country fixed effects (Model 1, Appendix \Cref{table:neg_sentiment_fixed}) and sending country fixed effects (Model 4, Appendix \Cref{table:neg_sentiment_fixed}). While the overall pattern is the same, the difference in the proportion of negativity in replies is even smaller in models with controls. Women receive on average from 0.018 to 0.25 percent points less negativity than men, depending on the model specification. The difference is no longer statistically significant when adding a binary control variable for whether the ambassadors themselves post above the median proportion of negative tweets or not (Model 3 and 6 in Appendix \Cref{table:neg_sentiment_fixed}). 

We validate the analysis by reiterating the models above with the proportion of positive replies as the dependent variable. The results point in the same direction. Women receive on average 6.1 percent points higher proportion of positive replies than men when running a simple OLS without controls ($p<0.01$). The pattern holds when including the control variables, as illustrated in \Cref{fig:dot_wiskers_sentiment}. Here, the difference ranges from  4.5 to 5.6 percent points (Model 1 and 4 in Appendix \Cref{table:pos_sentiment_fixed}). This is not surprising when taking into account that women post fewer negative tweets themselves (see Appendix \Cref{fig:amb_sentiment} and \Cref{table:amb_descriptives}) and that negative ambassador tweets are associated with a lower proportion of positive tone in the replies in our data. The difference between men and women ambassadors is no longer statistically significant when controlling for the proportion of negative tweets posted by the ambassadors and their country of origin through sending-country fixed effects. 

Is the correlation between gender and tone in the replies mitigated by the visibility or tone in the original tweets posted by the ambassadors themselves? We investigate this by testing Hypotheses 2.1 and 2.2.

\paragraph{Negativity and Visibility}
Hypothesis 2.1 predicts that the gender bias against women is stronger among ambassadors with high visibility, while Hypothesis 2.2 predicts that the bias increases among ambassadors who write more negative tweets themselves. In the case of Hypothesis H2.1, we would expect a positive interaction between gender and the number of retweets received by the ambassador, denoted by \emph{Woman $\times $ Above median n retweets}, when examining the proportion of negative tweets as the dependent variable. In other words, women with many retweets would receive more negativity than those with less visibility. Contrary to the hypothesis, the interaction term is \emph{negative}, statistically insignificant as well as both substantively small as shown in \Cref{fig:dot_wiskers_sentiment} and Appendix \Cref{table:neg_sentiment_fixed}, when controlling for number of original ambassador tweets, receiving country in-degree and retweets.

\paragraph{Negativity and Self-Negativity}
In cases of Hypothesis 2.2, we would observe a positive interaction term for: \emph{Woman $\times $ above median prop. negative ambassador tweets}, in the same figure and table. To put it differently, we would expect that women who go against the gender-stereotypical role by posting a higher proportion of negative tweets themselves would receive more negativity than women, who post fewer negative tweets. As shown in \Cref{fig:dot_wiskers_sentiment} and Appendix \Cref{table:neg_sentiment_fixed}, our results do not support this hypothesis: the interaction term is negative, substantively small, and statistically insignificant when using the same control variables as above. \

In addition, we find no evidence for Hypotheses 2.1. and 2.2 when replacing the dependent variable with the proportion of positive replies in the validation step (see  \Cref{fig:dot_wiskers_sentiment}). The results remain robust despite different model specifications: when using receiving country fixed effects, sending country fixed effects ( Appendix \Cref{table:neg_sentiment_fixed} and \Cref{table:pos_sentiment_fixed}), simple OLS only with variables of interest (Appendix \Cref{table:simple_OLS}). In sum, our data does not indicate that women receive more negative, public replies on a global level or that the potential gender bias through negative sentiment is mediated through visibility or the sentiment in the original ambassador tweets.

\subsection{Gendered Language}
 We now turn to examining whether women ambassadors are targeted with gendered language, as predicted by Hypothesis 3. In the first part of this section, we will examine the levels of dominance in the replies sent to the ambassadors. In the second part of the section, we will proceed to a more qualitative interpretation of the words associated with replies sent to men and women. 

 \subsubsection{Dominant Language}
This part of the analysis is based on the 1,367 ambassadors that have received at least one reply with a classified dominance score.\footnote{The number of ambassadors is lower in this part of the analysis because we were not able to infer dominance scores for 26.2\% of all of the tweets with known sentiment. This is due to the technological challenges of computing the complex measure for 65 languages. 57 of the 1,424 ambassadors received replies with inferred sentiment and not tweets with inferred dominance scores. This is equivalent to only 4\% of the full dataset.} The descriptive distribution is illustrated in \Cref{fig:mbassador_neg_dom_ret} C. The mean dominance score in replies sent to women is 0.013 higher compared to replies sent to men ($p<0.01$) when running an OLS model with no controls. The mean dominance score is 0.331 for men and 0.344 for women. The average dominance score is 3.9\% higher in replies sent to women than the average dominance score in the replies sent to men according to these estimates. 
 
 As shown in Appendix \Cref{table:dominance_fixed}, the average dominance score in replies sent to women ranges from 0.010 to 0.012 higher than that sent to men -- depending on the model specification. The difference remains statistically significant even when controlling for the number of ambassador tweets, visibility through retweets, the prestige of the country that the ambassadors are assigned to (measured as in-degree), and the receiving or sending country through fixed effects. In other words, the average level of dominance in the replies sent to women is at least 3.1\% higher than the levels of dominance sent to men -- according to the most conservative estimate, when including the controls. In an additional analysis outside of this section, we find no evidence suggesting that women with higher visibility (measured as retweets) receive more dominance in replies than women with fewer retweets. The same pattern holds when comparing women with above median proportion of negative original ambassador tweets with women who post fewer negative tweets.

Overall, our analysis shows that the language used by Twitter users who reply to women and men ambassadors is indeed gendered in line with Hypothesis 3. The difference is not high, but it is important when considering that ambassadors collectively are responded to by millions of tweets throughout multiple years. Notably, the global difference is persistent when incorporating approximately 65 languages sent to 1,367 ambassadors from a total of 148 countries. 

\subsubsection{Lexical Biases}
In this section, we investigate the lexical differences in responses to men and women ambassadors. To this end, we analyze word usage towards ambassadors being received by countries with the highest number of tweets in response. The top 10 women- and men-associated words identified using $\pmi$ for the top 5 countries receiving ambassadors with the highest number of responses are shown in \Cref{tab:pmi}.

\begin{table}[H]
\centering
\fontsize{10}{10}\selectfont
\begin{tabular}{@{}ll@{}}
\toprule
Receiving country & Associated words \\ \midrule
\multicolumn{2}{l}{Men-biased} \\ \midrule
India   & sir, support, Israel, love, friend, students, open, India, visa, decision\\
Brazil    & China, party, Brazil, people, thank, help, way, country, years, respect\\
United States &  Tigray, tigraygenocide, Ethiopia, Chinese, lie, ethnic, Ethiopian, Irish, rape, \\
& independent \\
Lebanon     &  Mr, Saudi, government, new, come, send, times, company, appreciate, big \\
Iraq     &  Iraq, amp, militias, hope, time, UK, Iraqi, government, people, country\\ \midrule
\multicolumn{2}{l}{Women-biased} \\\midrule
India   & Finland, Finnish, engage, software, actively, cheated, flowers, owners, heargaza, plus \\
Brazil  & happy, culture, brain, time, technology, terrorism, want, know, ministers, best \\
United States  &  Saudi, salmon, mbs, saudiarabia, highness, Arabia, colored, prince, Arab, queens \\
Lebanon  & teachers, quality, general, UNRWA, decision, Australian, jobs, Gaza, paid, learn \\
Iraq  &  president, bless, Jordan, national, office, interesting, kdp, missions, official, asking \\ \bottomrule
\end{tabular}
\caption[Men vs women $\pmi$ results]{The top-10 men (top) and women-biased (bottom) words in the dataset for the top-5 receiving countries with the highest numbers of tweets written in response to the ambassadors, using $\pmi$.}
\label{tab:pmi}
\end{table}

We observe that the words written in response to ambassadors whether men or women cover predominantly international politics and diplomacy. We verify this by employing Latent Dirichlet analysis (LDA; \citealt{Blei2003LDA}) on the subset of tweets in English written in response to the ambassadors. We find that indeed the topics covered evolve to a high degree around politics and diplomacy (see \Cref{tab:topic} in the Appendix). Additional validation of both results related to negativity and gendered language is available in the Validation section in Appendix C.





\section{Conclusion}
\label{sec:conclusion}
Historically, diplomacy has been a men-dominated field, with women facing various forms of discrimination and bias. While there have been changes, gender inequalities persist in the diplomatic profession. This study's focus on social media, specifically Twitter, is significant as Twitter has become the platform of choice for diplomats worldwide, making it a critical space for shaping international discourse.

Our findings challenge common assumptions about online gender bias against women. Contrary to expectations, women ambassadors do not face a higher degree of outright negativity in responses to their tweets on a global scale. Although they do receive more gendered language in the replies, the difference is not substantively large. Instead, the primary source of online bias against women ambassadors is their lower online visibility. This subtler form of bias, while less overt, is of paramount importance, as it affects their ability to engage in public diplomacy effectively. We also observe an interesting online glass-ceiling effect: The relatively few women who gain prestigious, diplomatic positions in the men-dominated diplomatic sphere experience an additional barrier at the top in the competition for online visibility. The implications of these findings are twofold. On one hand, the diplomatic arena may provide a relatively `safer' online space for women compared to other political domains. On the other hand, it underscores the deeply ingrained nature of these biases, including the value attached to tweets by women ambassadors compared to their men colleagues.

By conducting the first global-scale, systematic study of gender bias in digital diplomacy, the research not only sheds light on the multifaceted nature of online gender bias but also provides essential methodological insights for future investigations and a foundation for cross-temporal and cross-platform comparisons. Our findings illustrate the methodological importance of combining analysis of publicly available content with a more network-centered analysis of retweets. While gender bias may appear small in the publicly available content, one risks overlooking inequality in a more latent, yet highly important resource, online visibility itself. The findings illuminate subtle biases in interactions between ambassadors and their audiences, leaving a more detailed analysis of how ambassadors interact with other ambassadors outside of the scope of this study. 

Our research not only advances our understanding of gender bias in digital diplomacy but also contributes to the broader conversation on gender equality in international politics. It underscores the importance of exploring how to ensure greater online visibility for women ambassadors, as this visibility is not just a matter of representation but also a fundamental resource for engaging in diplomatic activities on social media. As such, this study provides a critical foundation for future research and policy development aimed at reducing gender bias in the realm of digital diplomacy and beyond.

The study is limited to Twitter, which is the main home of digital diplomacy even today. However, gender bias patterns may be different when looking at other social media, and perhaps even the future versions of the same platform. Social media platforms constantly change due to updated algorithms, moderation policies, and functions. As ``Twitter'' has transitioned into ``X'', the information environment on Twitter is changing, potentially also the gender bias we observe. 

Lastly, the study does not measure how gender interacts with other biases related to the ambassadors' ethnicity, perceived race, religion, or other factors. We hope that our multilingual and globally-spanning study will serve as a stepping stone for future research on intersectionality in digital diplomacy. 



\section{Acknowledgements}
This research was co-funded by the European Research Council as part of the StG DIPLOFACE 680102, a DFF Research Project 1 under grant agreement No 9130-00092B, and supported by the Pioneer Centre for AI, DNRF grant number P1.

\newpage
\bibliography{references,references_2023_manual, bias_nlp}

\newpage
\renewcommand{\thefigure}{A\arabic{figure}}
\setcounter{figure}{0}
\renewcommand{\thetable}{A\arabic{table}}
\setcounter{table}{0}

\section{Appendix}

\

\subsection*{Appendix A: Descriptive Statistics}
\renewcommand{\thefigure}{A\arabic{figure}}
\setcounter{figure}{0}
\renewcommand{\thetable}{A\arabic{table}}
\setcounter{table}{0}

 \begin{figure}[H]
\centering
\caption{The figure shows the aggregated distribution of negative, positive and neutral replies sent to ambassadors in their destination countries. Vertical lines reflect means. Values in the lowest row reflect the difference between the mean proportion of sentiment (positive, negative or neutral) for men minus the mean proportion for women. }
\includegraphics[width=0.95\textwidth]{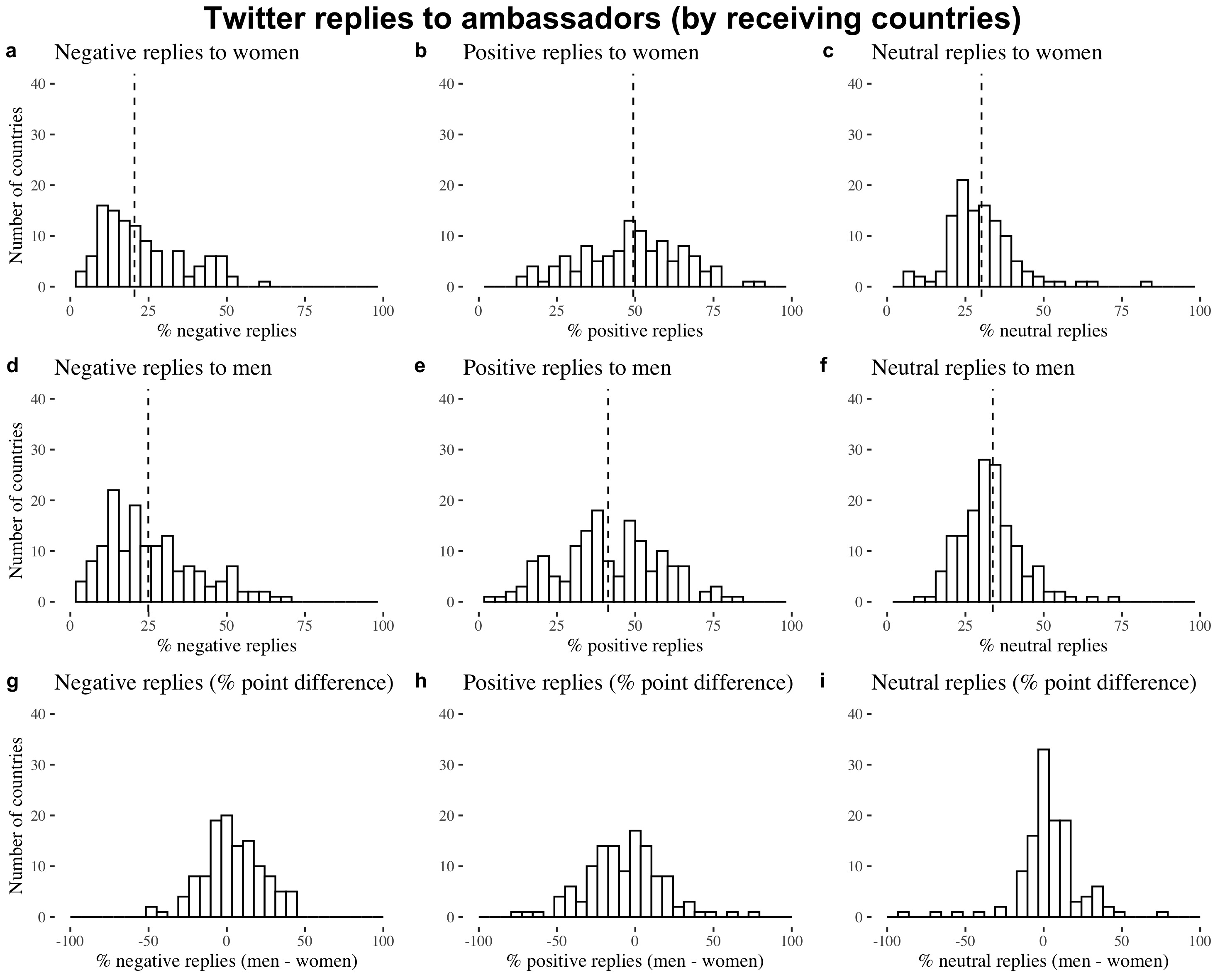}

\label{fig:his_replies}
\end{figure}

 \begin{figure}[H]
\centering
\caption{Node size reflects the number of replies. Columns to the left reflect values for ambassadors who are \emph{sent to} respective destinations, while columns to the right reflect values for ambassadors originating \emph{from} the given countries.}
\includegraphics[width=0.95\textwidth]{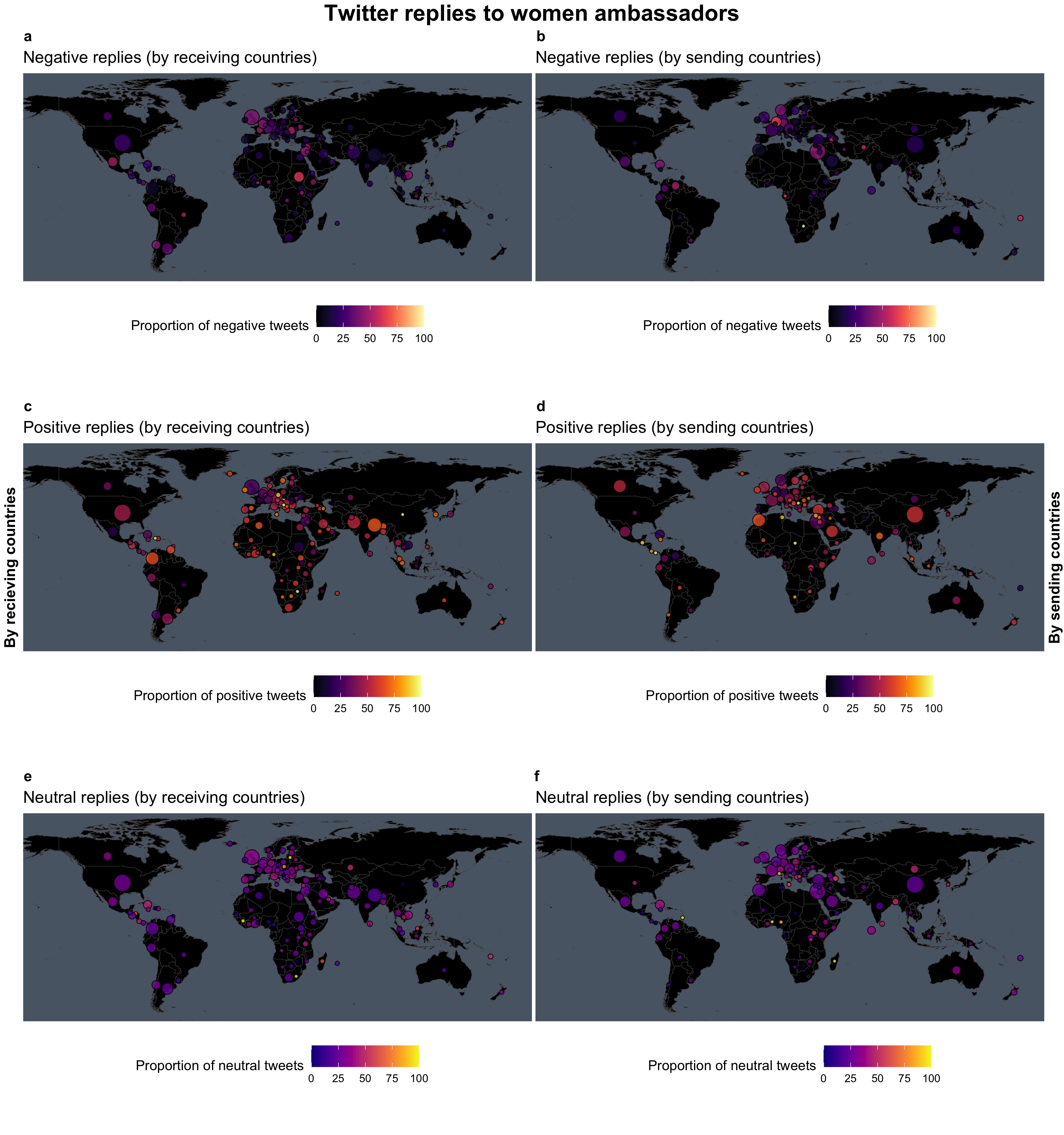}

\label{fig:women_map_sent}
\end{figure}

 \begin{figure}[H]
\centering
\caption{Node size reflects number of replies. Grey colour indicates that there are only men in the subsample. Colour reflects the difference between the mean proportion of sentiment (positive, negative, or neutral) for men minus women. For example, lower (darker) values in Figure \textbf{a} indicate that women ambassadors on average receive a higher proportion of negative replies than their men colleagues who are sent to the respective country.}
\includegraphics[width=0.95\textwidth]{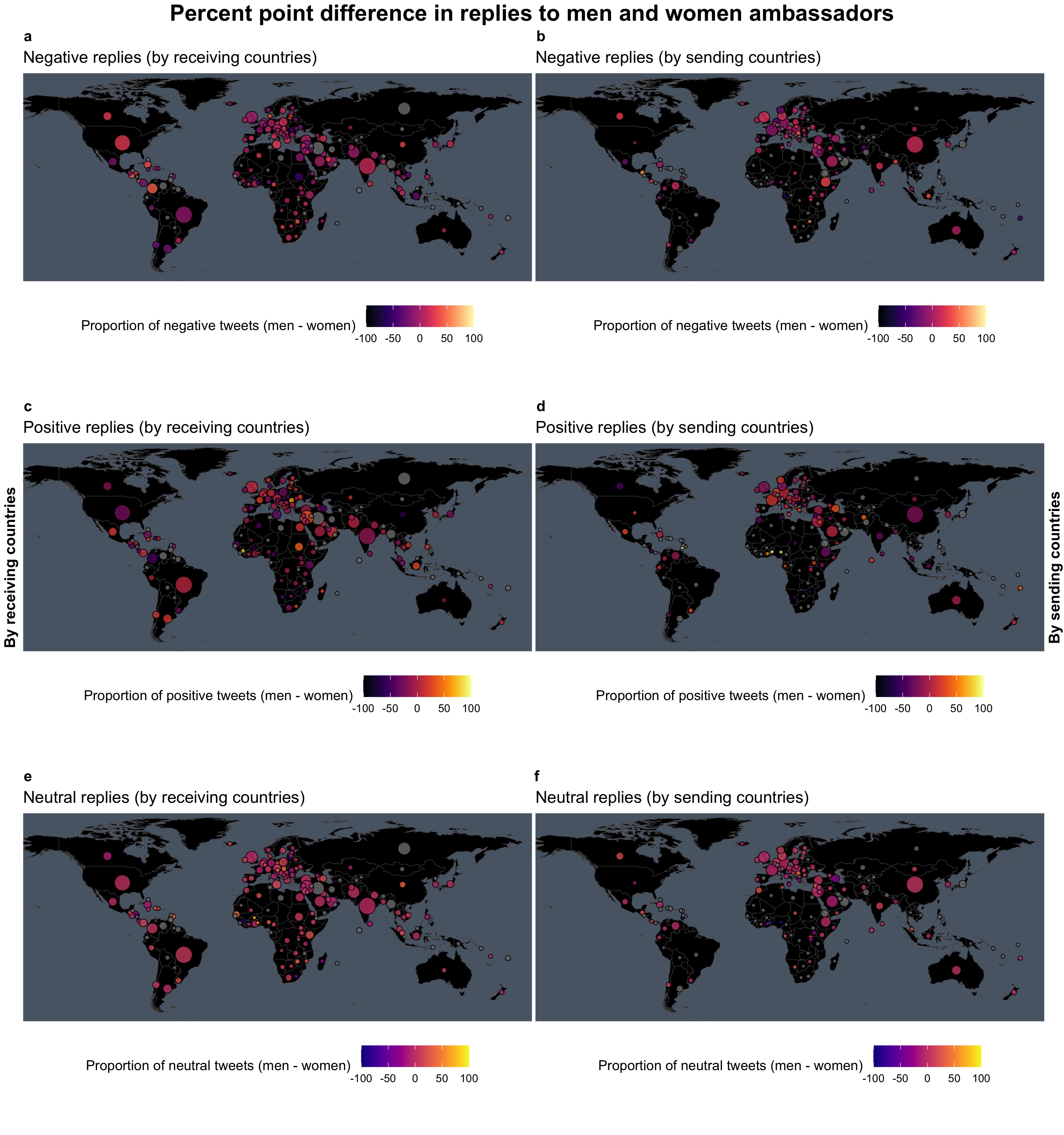}

\label{fig:men_minus:map_sent}
\end{figure}


\begin{figure}[H]

 \caption{Sentiment in tweets posted by ambassadors (density plots)}
\centering
\includegraphics[width=0.9\textwidth]{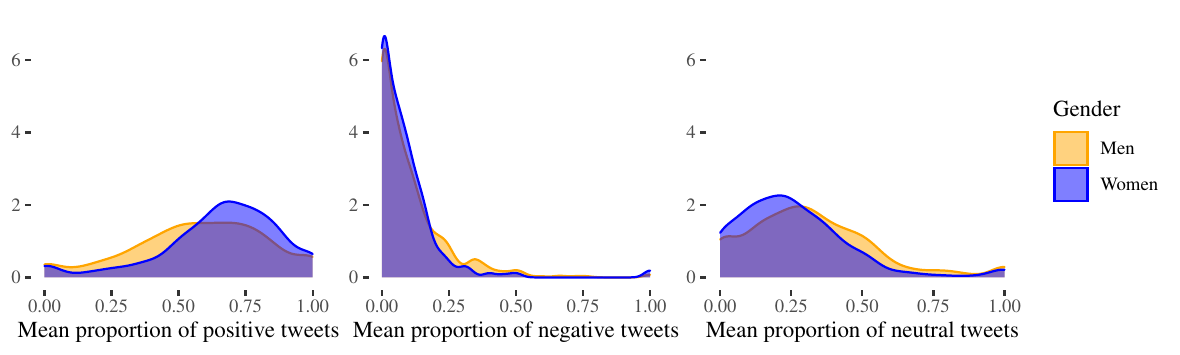}

\label{fig:amb_sentiment}
\end{figure}

\begin{figure}[H]
\label{fig:amb_VAD}
 \caption{Valence, arousal, and dominance scores in tweets posted by ambassadors (density plots)}
\centering
\includegraphics[width=0.9\textwidth]{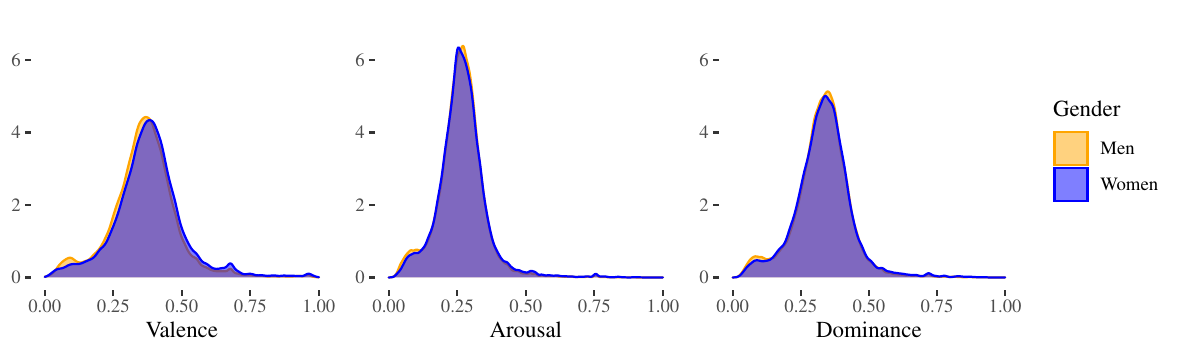}

\end{figure}

\begin{figure}[H]
\label{fig:mean_amb_VAD}
 \caption{Mean valence, arousal and dominance scores (for each ambassador) in tweets posted by ambassadors (density plots)}
\centering
\includegraphics[width=0.9\textwidth]{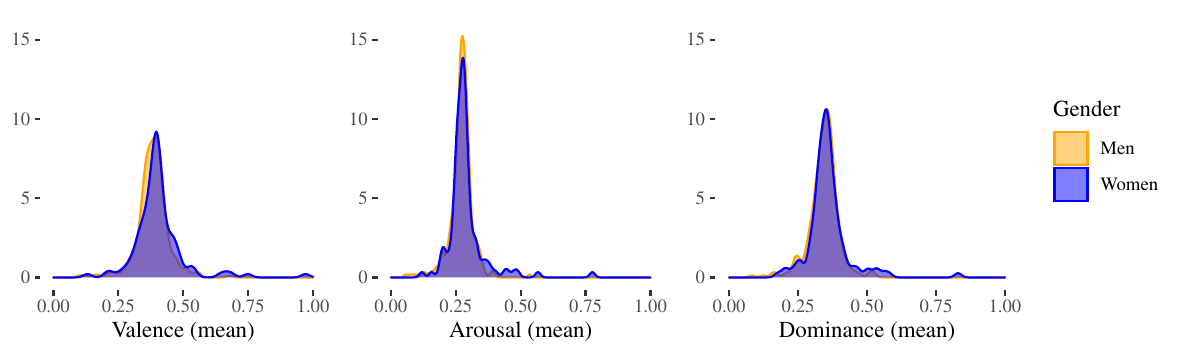}

\end{figure}

\begin{table}[ht]

\caption{Descriptive overview of sentiment, valence, arousal, and dominance scores in tweets posted by ambassadors}
\centering
\begin{tabular}{rll}
  \hline
  \hline
 & Men & Women \\ 
   \hline
  Total number of tweets & 104,531 &  37,012 \\ 
  Mean tweets & 91.94 & 78.58 \\ 
  SD tweets & 150.45 & 153.00 \\ 
  Median tweets & 42 & 39 \\ 
  Mean \% of positive tweets & 0.58 & 0.65 \\ 
  SD \% of positive tweets & 0.25 & 0.23 \\ 
  Mean \% of negative tweets & 0.10 & 0.09 \\ 
  SD \% of negative tweets & 0.14 & 0.13 \\ 
  Mean \% of neutral tweets & 0.32 & 0.26 \\ 
  SD \% of neutral tweets & 0.23 & 0.20 \\ 
  Mean Valence & 0.36 & 0.38 \\ 
  SD Valence & 0.12 & 0.13 \\ 
  Mean Arousal & 0.26 & 0.27 \\ 
  SD Arousal & 0.08 & 0.09 \\ 
  Mean Dominance & 0.33 & 0.33 \\ 
  SD Dominance & 0.10 & 0.10 \\ 
   \hline
\end{tabular}
\label{table:amb_descriptives}
\end{table}

\addcontentsline{toc}{subsection}{Appendix B: Regression models}
\subsection*{Appendix B: Regression Models}
\renewcommand{\thefigure}{B\arabic{figure}}
\setcounter{figure}{0}
\renewcommand{\thetable}{B\arabic{table}}
\setcounter{table}{0}

\begin{table}[H]
\caption{Negative Binomial Multilevel models: Estimating difference in visibility measured as retweet and follower count. Ambassadors are nested in their respective receiving countries (Models 1 and 3) and sending countries (Models 2 and 4). The coefficients reflect log change in the dependent variables per unit change in the independent variables.}

\begin{center}
\begin{tabular}{l c c c c}
\hline \\[-1.8ex] 
 & \multicolumn{4}{c}{\textit{Dependent variables:}} \\

\\[-1.8ex] & \multicolumn{2}{c}{Retweets} & \multicolumn{2}{c}{Followers} \\ 
 & Model 1 & Model 2 & Model 3 & Model 4 \\
\hline
Intercept                       & $-0.90^{***}$ & $-1.20^{***}$ & $6.26^{***}$ & $5.73^{***}$ \\
                                & $(0.20)$      & $(0.18)$      & $(0.16)$     & $(0.14)$     \\
in-degree (receiving country)    & $1.65^{***}$  & $1.16^{***}$  & $1.60^{***}$ & $1.66^{***}$ \\
                                & $(0.42)$      & $(0.18)$      & $(0.37)$     & $(0.15)$     \\
Woman                           & $-0.61^{***}$ & $-0.45^{***}$ & $-0.18^{*}$  & $-0.19^{*}$  \\
                                & $(0.10)$      & $(0.09)$      & $(0.08)$     & $(0.08)$     \\
log(n tweets + 0.1)             & $1.17^{***}$  & $1.17^{***}$  & $0.36^{***}$ & $0.31^{***}$ \\
                                & $(0.02)$      & $(0.03)$      & $(0.01)$     & $(0.01)$     \\
\hline
AIC                             & $18636.63$    & $18333.98$    & $34977.08$   & $34599.22$   \\
Log Likelihood                  & $-9312.31$    & $-9160.99$    & $-17482.54$  & $-17293.61$  \\
Num. obs.                       & $1960$        & $1960$        & $1948$       & $1948$       \\
Num. groups: Receiving country     & $172$         &             & $172$        &            \\
Var: Receiving country (Intercept) & $0.90$        &             & $0.70$       &            \\
Num. groups: Sending country       &             & $164$         &            & $163$        \\
Var: Sending country (Intercept)   &             & $1.41$        &            & $1.50$       \\
\hline
\multicolumn{5}{l}{\scriptsize{$^{***}p<0.001$; $^{**}p<0.01$; $^{*}p<0.05$}}
\end{tabular}
\label{table:reg_vis_multi}
\end{center}
\end{table}

\begin{table}[H]
\caption{Negative Binomial models with country-level fixed effects: Estimating difference in visibility measured as retweet and follower count}
 \resizebox{\linewidth}{!}{%
\begin{tabular}{lcccc}
\tabularnewline\midrule\midrule
Dependent Variables:&\multicolumn{2}{c}{Retweets}&\multicolumn{2}{c}{Followers}\\
Model:&(1) & (2) & (3) & (4)\\
\midrule \emph{Variables}&   &   &   &  \\
Woman&-0.4022$^{***}$ & -0.5538$^{***}$ & -0.1732 & -0.1201\\
  &(0.1341) & (0.1348) & (0.1141) & (0.1223)\\
log(n tweets+0.1)&1.155$^{***}$ & 1.199$^{***}$ & 0.3018$^{***}$ & 0.3694$^{***}$\\
  &(0.0423) & (0.0687) & (0.0269) & (0.0277)\\
in-degree (receiving country)&1.213$^{***}$ &    & 1.711$^{***}$ &   \\
  &(0.3272) &    & (0.2338) &   \\
\midrule \emph{Fixed-effects}&   &   &   &  \\
Sending country& Yes &  & Yes & \\
Receiving country &  & Yes &  & Yes\\
\midrule \emph{Fit statistics}&  & & & \\
Standard-Errors& Sending country&Receiving country&Sending country&Receiving country\\
Squared Correlation & 0.13338&0.10530&0.14987&0.09290\\
BIC & 19,003.7&19,518.6&35,362.8&35,883.9\\
Over-dispersion & 0.50944&0.44309&0.59918&0.50779\\
\midrule\midrule\multicolumn{5}{l}{\emph{Signif. Codes: ***: 0.01, **: 0.05, *: 0.1}}\\
\end{tabular}}
\label{table:reg_vis_multi_fixed}
\end{table}

\begin{table}[H]
\caption{OLS with country-level fixed effects: The difference in proportion of negative replies sent to the ambassadors}
 \resizebox{\linewidth}{!}{%
\begin{tabular}{lcccccc}
\tabularnewline\midrule\midrule
Dependent Variable:&\multicolumn{6}{c}{Proportion of negative replies}\\
Model:&(1) & (2) & (3) & (4) & (5) & (6)\\
\midrule \emph{Variables}&   &   &   &   &   &  \\
Woman&-0.0244$^{**}$ & -0.0163 & -0.0030 & -0.0182$^{*}$ & -0.0079 & 0.0007\\
  &(0.0100) & (0.0171) & (0.0143) & (0.0106) & (0.0174) & (0.0154)\\
log(n tweets +0.1)&-0.0014 & -0.0015 & -0.0086$^{*}$ & -0.0064 & -0.0066 & -0.0121$^{**}$\\
  &(0.0049) & (0.0049) & (0.0047) & (0.0055) & (0.0055) & (0.0056)\\
Above median n retweets&0.0785$^{***}$ & 0.0834$^{***}$ & 0.0837$^{***}$ & 0.0503$^{***}$ & 0.0569$^{***}$ & 0.0559$^{***}$\\
  &(0.0115) & (0.0147) & (0.0114) & (0.0163) & (0.0161) & (0.0157)\\
Woman $\times $ Above median n retweets&   & -0.0174 &    &    & -0.0215 &   \\
  &   & (0.0239) &    &    & (0.0228) &   \\
Above median prop. negative ambassador tweets&   &    & 0.0932$^{***}$ &    &    & 0.0930$^{***}$\\
  &   &    & (0.0122) &    &    & (0.0110)\\
Woman $\times $ above median prop. negative ambassador tweets&   &    & -0.0327 &    &    & -0.0290\\
  &   &    & (0.0234) &    &    & (0.0211)\\
in-degree (receiving)&   &    &    & 0.0667$^{***}$ & 0.0668$^{***}$ & 0.0717$^{***}$\\
  &   &    &    & (0.0225) & (0.0225) & (0.0221)\\
\midrule \emph{Fixed-effects}&   &   &   &   &   &  \\
Receiving country & Yes & Yes & Yes &  &  & \\
Sending country &  &  &  & Yes & Yes & Yes\\
\midrule \emph{Fit statistics}&  & & & & & \\
Standard-Errors& Receiving country&Receiving country&Receiving country&Sending country&Sending country&Sending country\\
Observations & 1,424&1,424&1,424&1,424&1,424&1,424\\
\midrule\midrule\multicolumn{7}{l}{\emph{Signif. Codes: ***: 0.01, **: 0.05, *: 0.1}}\\
\end{tabular}}
\label{table:neg_sentiment_fixed}
\end{table}

\begin{table}[H]
\caption{OLS with country-level fixed effects: The difference in the proportion of positive replies sent to the ambassadors}
 \resizebox{\linewidth}{!}{%
\begin{tabular}{lcccccc}
\tabularnewline\midrule\midrule
Dependent Variable:&\multicolumn{6}{c}{Proportion of positive replies}\\
Model:&(1) & (2) & (3) & (4) & (5) & (6)\\
\midrule \emph{Variables}&   &   &   &   &   &  \\
Woman&0.0569$^{***}$ & 0.0538$^{*}$ & 0.0457$^{*}$ & 0.0401$^{**}$ & 0.0408 & 0.0305\\
  &(0.0190) & (0.0313) & (0.0248) & (0.0165) & (0.0273) & (0.0222)\\
log(n tweets +0.1)0.0221$^{***}$ & 0.0222$^{***}$ & 0.0297$^{***}$ & 0.0286$^{***}$ & 0.0286$^{***}$ & 0.0326$^{***}$\\
  &(0.0057) & (0.0057) & (0.0057) & (0.0054) & (0.0054) & (0.0053)\\
Above median n retweets&-0.0596$^{***}$ & -0.0614$^{***}$ & -0.0654$^{***}$ & -0.0243 & -0.0238 & -0.0283\\
  &(0.0174) & (0.0209) & (0.0175) & (0.0198) & (0.0213) & (0.0196)\\
Woman $\times $ above median n retweets&   & 0.0066 &    &    & -0.0015 &   \\
  &   & (0.0363) &    &    & (0.0299) &   \\
Above median prop. negative ambassador tweets&   &    & -0.0925$^{***}$ &    &    & -0.0645$^{***}$\\
  &   &    & (0.0163) &    &    & (0.0196)\\
Woman $\times $ above median prop. negative ambassador tweets&   &    & 0.0107 &    &    & 0.0124\\
  &   &    & (0.0344) &    &    & (0.0313)\\
in-degree (receiving)&   &    &    & -0.0378 & -0.0378 & -0.0409\\
  &   &    &    & (0.0298) & (0.0298) & (0.0302)\\
\midrule \emph{Fixed-effects}&   &   &   &   &   &  \\
Receiving country & Yes & Yes & Yes &  &  & \\
Sending country &  &  &  & Yes & Yes & Yes\\
\midrule \emph{Fit statistics}&  & & & & & \\
Standard-Errors& Receiving country&Receiving country&Receiving country&Sending country&Sending country&Sending country\\
Observations & 1,424&1,424&1,424&1,424&1,424&1,424\\
\midrule\midrule\multicolumn{7}{l}{\emph{Signif. Codes: ***: 0.01, **: 0.05, *: 0.1}}\\
\end{tabular}}
\label{table:pos_sentiment_fixed}
\end{table}

\begin{table}[H]
\caption{OLS with country-level fixed effects: The difference in average dominance scores in replies sent to the ambassadors}
 \resizebox{\linewidth}{!}{%
\begin{tabular}{lcccc}
\tabularnewline\midrule\midrule
Dependent Variable:&\multicolumn{4}{c}{Mean Dominance score}\\
Model:&(1) & (2) & (3) & (4)\\
\midrule \emph{Variables}&   &   &   &  \\
Woman&0.0116$^{***}$ & 0.0110$^{***}$ & 0.0101$^{***}$ & 0.0102$^{***}$\\
  &(0.0038) & (0.0036) & (0.0038) & (0.0036)\\
log(n tweets +0.1)&   &    & 0.0031$^{**}$ & 0.0037$^{**}$\\
  &   &    & (0.0014) & (0.0014)\\
Above median n retweets&   &    & -0.0178$^{***}$ & -0.0079\\
  &   &    & (0.0053) & (0.0050)\\
in-degree (receiving)&   &    &    & -0.0228$^{***}$\\
  &   &    &    & (0.0079)\\
\midrule \emph{Fixed-effects}&   &   &   &  \\
Receiving country & Yes &  & Yes & \\
Sending country &  & Yes &  & Yes\\
\midrule \emph{Fit statistics}&  & & & \\
Standard-Errors& Receiving country&Sending country& Receiving country&Sending country\\
Observations & 1,367&1,367&1,367&1,367\\
\midrule\midrule\multicolumn{5}{l}{\emph{Signif. Codes: ***: 0.01, **: 0.05, *: 0.1}}\\
\end{tabular}}
\label{table:dominance_fixed}
\end{table}

\newpage

\begin{table}[!htbp] \centering 
  \caption{Simple OLS only with variables of interest: The difference in the proportion of negative and positive
replies sent to the ambassadors} 
  \label{table:simple_OLS} 
     \resizebox{\linewidth}{!}{%
\begin{tabular}{@{\extracolsep{5pt}}lcccccc} 
\\[-1.8ex]\hline 
\hline \\[-1.8ex] 
 & \multicolumn{6}{c}{\textit{Dependent variable:}} \\ 
\cline{2-7} 
\\[-1.8ex] & \multicolumn{3}{c}{Proportion of negative replies} & \multicolumn{3}{c}{Proportion of positive replies} \\ 
\\[-1.8ex] & (1) & (2) & (3) & (4) & (5) & (6)\\ 
\hline \\[-1.8ex] 
 Woman & $-$0.037$^{***}$ & $-$0.020 & $-$0.014 & 0.061$^{***}$ & 0.060$^{***}$ & 0.049$^{**}$ \\ 
  & (0.012) & (0.016) & (0.016) & (0.016) & (0.022) & (0.023) \\ 
  & & & & & & \\ 
 Above median n retweets &  & 0.077$^{***}$ &  &  & $-$0.005 &  \\ 
  &  & (0.013) &  &  & (0.018) &  \\ 
  & & & & & & \\ 
 Woman $\times $ above median n retweets &  & $-$0.025 &  &  & 0.001 &  \\ 
  &  & (0.024) &  &  & (0.033) &  \\ 
  & & & & & & \\ 
 Above median prop. negative ambassador tweets &  &  & 0.104$^{***}$ &  &  & $-$0.071$^{***}$ \\ 
  &  &  & (0.012) &  &  & (0.017) \\ 
  & & & & & & \\ 
 Woman $\times $ above median prop. negative ambassador tweets &  &  & $-$0.037 &  &  & 0.017 \\ 
  &  &  & (0.023) &  &  & (0.033) \\ 
  & & & & & & \\ 
 Constant & 0.209$^{***}$ & 0.168$^{***}$ & 0.155$^{***}$ & 0.436$^{***}$ & 0.439$^{***}$ & 0.473$^{***}$ \\ 
  & (0.006) & (0.009) & (0.009) & (0.009) & (0.013) & (0.013) \\ 
  & & & & & & \\ 
\hline \\[-1.8ex] 
Observations & 1,424 & 1,424 & 1,424 & 1,424 & 1,424 & 1,424 \\ 
R$^{2}$ & 0.007 & 0.037 & 0.061 & 0.010 & 0.010 & 0.024 \\ 
Adjusted R$^{2}$ & 0.006 & 0.035 & 0.059 & 0.009 & 0.008 & 0.021 \\ 
\hline 
\hline \\[-1.8ex] 
\textit{Note:}  & \multicolumn{6}{r}{$^{*}$p$<$0.1; $^{**}$p$<$0.05; $^{***}$p$<$0.01} \\ 
\end{tabular}} 
\end{table} 

\newpage
\addcontentsline{toc}{subsection}{Appendix C: Validation}
\subsection*{Appendix C: Validation}
\renewcommand{\thefigure}{C\arabic{figure}}
\setcounter{figure}{0}
\renewcommand{\thetable}{C\arabic{table}}
\setcounter{table}{0}

\begin{figure}[H]
\centering
\caption{}
\includegraphics[width=0.9\textwidth]{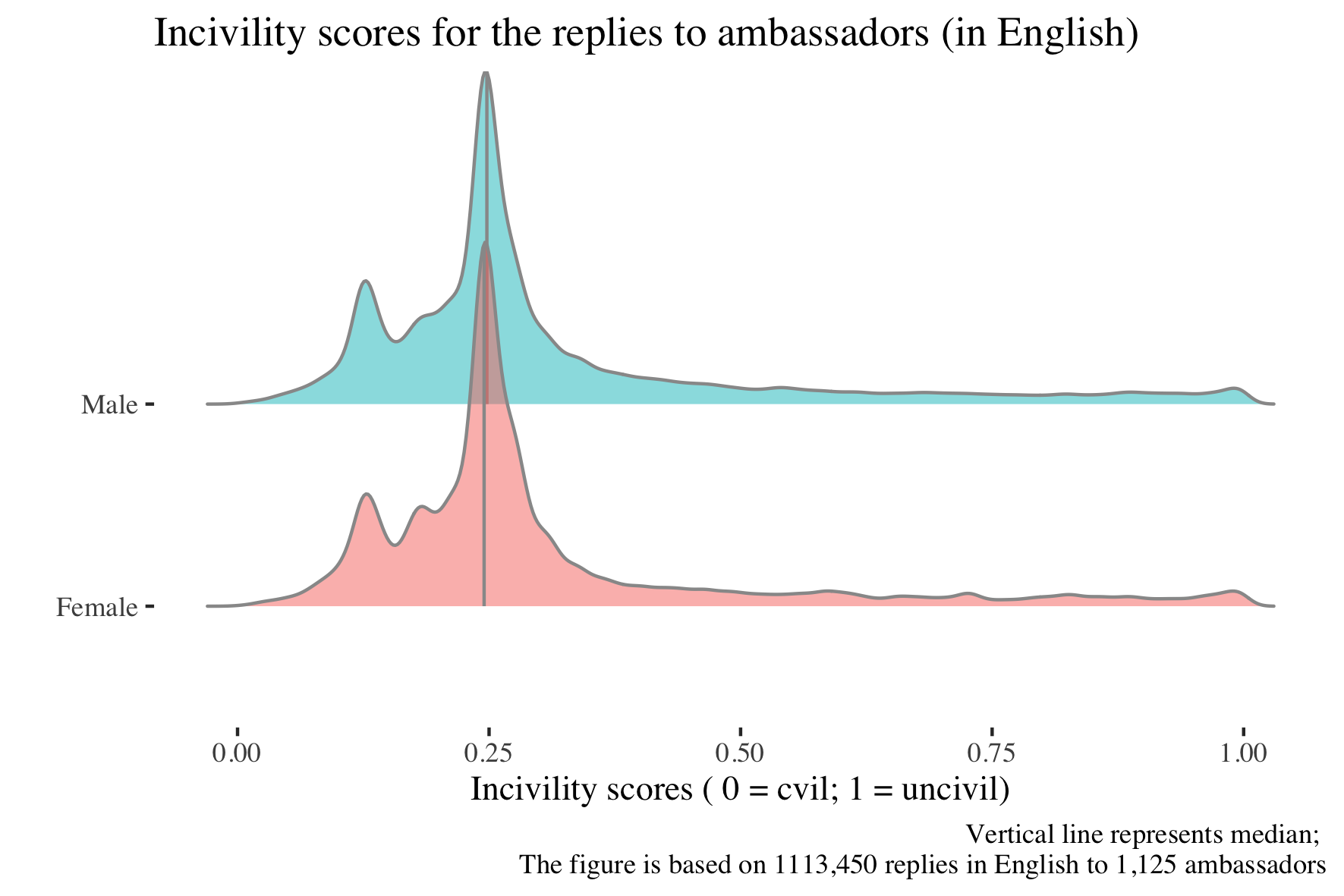}

\label{fig:amb_incivil}
\end{figure}

We validate the results through three steps. 
In the first step, we validate the findings on negativity by detecting incivility in direct replies sent to ambassadors in English. For this purpose, we use a machine-learning-based incivility algorithm originally developed by \citet{theocharis_dynamics_2020} for their study of tweets sent to US members of Congress. The score ranges from 0 (civil) to 1 (uncivil). We find that women receive on average slightly more civil tweets than their men colleagues. This further corroborates the results based on sentiment, which indicate that women do not receive more negatively than men \emph{overall}. The density plot below gives a descriptive overview of the distribution of the incivility score (\Cref{fig:amb_incivil}).

In the second step, we verify that replies sent to women ambassadors are mainly related to political topics by using Latent Dirichlet analysis (LDA; \citealt{Blei2003LDA}) to identify topics in direct replies sent to ambassadors in English. We find that indeed the topics covered evolve to a high degree around politics and diplomacy (see \Cref{tab:topic} in the Appendix). 

\begin{table}[H]
\centering
\fontsize{10}{10}\selectfont
\begin{tabular}{@{}ll@{}}
\toprule
Topic & Associated words \\ \midrule
Topic 1 & China, annlinde, Tigray, world, Chinese, people, time, act, EU, take \\
Topic 2 & annlinde, Pakistan, country, Canada, world, people, women, one, see, like \\
Topic 3 & Israel, people, children, terrorist, crimes, Israeli, Palestinian, state, Palestine \\
Topic 4 & you, women, ..., marisepayne, what, Australia, n't, like, get \\
Topic 5 & Myanmar, military, please, people, junta, election, respect, whatshappeninginmyanmar, \\ 
& ASEAN, coup \\
Topic 6 & human, happy, rights, annlinde, you, new, this, day, peak, please \\
Topic 7 & thank, thanks, great, help, good, much, ambassador, excellency, congratulations \\
Topic 8 & want, respect, government, !!!, need, !!, leader, votes, another, please \\
Topic 9 & not, minister, foreign, meet, please, people, n't, represent, Myanmar, Thailand \\

 \bottomrule
\end{tabular}
\caption[Topic Modelling]{The top-9 identified topics based on all tweets written in English in response to the ambassadors.}
\label{tab:topic}
\end{table}

In the last step, we examine whether gender bias through retweets, negativity and dominance in replies is mediated by the overall levels of gender inequality in the respective countries. To ensure robustness, we operationalize country-level gender inequality using three, established measures: 1) Gender Inequality Index \citep{GII_2020,QOG_2022}, Proportion of seats held by women in national parliaments \citep{World_Bank2021, QOG_2022} and Gender Social Norms Index \citep{GSNI_2020} with the latest observations as of 2019. These results as shown in the regression tables in Appendix D. The overall finding remains the same: We find a strong gender bias in visibility through retweets even when taking into account gender inequality in the sending or receiving country. In addition to this, we find no evidence that women receive fewer retweets, more negativity or dominant language in replies if they originate from- or are sent to countries with above-median levels of gender inequality. We observe one exception: Women who are sent to a country with a high Gender Social Norms Index receive slightly more dominant replies, however, the difference is not substantial.

\newpage
\addcontentsline{toc}{subsection}{Appendix D: Gender inequality in sending and receiving countries}
\subsection*{Appendix D: Gender Inequality in Sending and Receiving Countries}
\label{app:d}
\renewcommand{\thefigure}{D\arabic{figure}}
\setcounter{figure}{0}
\renewcommand{\thetable}{D\arabic{table}}
\setcounter{table}{0}
This appendix presents additional robustness checks. Models in Tables D1 to D6 test whether the difference in visibility between men and women is mediated by levels of gender inequality in the respective countries. The latter is operationalized as 1) Gender Inequality Index, 2) Proportion of women in parliament and 3) Gender Social Norms Index. The respective models take into account either the gender inequality in the \emph{sending country}, i.e., an ambassador's country of origin, or the \emph{receiving country} that the ambassador is assigned to. The number of observations varies because it is not possible to obtain the same inequality measurements for all countries. Models in Table D7 test whether the difference in visibility is mediated by the prestige of the receiving country, measured as that country's in-degree in a diplomatic network, where each tie reflects one country sending an embassy to another country. Models in Table D8 test whether negativity is mediated by either prestige (in-degree), Gender Inequality Index or proportion of women in parliament in the receiving country. 

\begin{table}[H]
\label{table: gii_rec}
\caption{Negative Binomial Multilevel models: Gender Inequality Index in receiving countries and visibility. Estimating difference in visibility measured as retweet and follower count. Ambassadors are nested in their respective receiving countries (Models 1 and 3) and sending countries (Models 2 and 4). The coefficients reflect log change in the dependent variables per unit change in the independent variables.}
\begin{center}
\begin{tabular}{l c c c c}
\tabularnewline\midrule\midrule
Dependent Variables:&\multicolumn{2}{c}{Retweets}&\multicolumn{2}{c}{Followers}\\
Model:&(1) & (2) & (3) & (4)\\
\hline
Intercept                       & $-1.73^{***}$ & $-1.87^{***}$ & $5.80^{***}$ & $5.50^{***}$ \\
                                & $(0.27)$      & $(0.20)$      & $(0.23)$     & $(0.16)$     \\
in-degree (receiving country)    & $2.53^{***}$  & $1.70^{***}$  & $1.96^{***}$ & $1.85^{***}$ \\
                                & $(0.42)$      & $(0.19)$      & $(0.40)$     & $(0.16)$     \\
Woman                           & $-0.62^{***}$ & $-0.36^{**}$  & $-0.06$      & $-0.16$      \\
                                & $(0.14)$      & $(0.12)$      & $(0.12)$     & $(0.11)$     \\
Gender Inequality Index         & $0.73^{***}$  & $0.63^{***}$  & $0.56^{**}$  & $0.29^{**}$  \\
                                & $(0.20)$      & $(0.10)$      & $(0.19)$     & $(0.09)$     \\
log(n tweets + 0.1)             & $1.18^{***}$  & $1.19^{***}$  & $0.36^{***}$ & $0.31^{***}$ \\
                                & $(0.02)$      & $(0.03)$      & $(0.01)$     & $(0.01)$     \\
Woman x Gender Inequality Index & $0.04$        & $-0.02$       & $-0.32$      & $-0.02$      \\
                                & $(0.20)$      & $(0.17)$      & $(0.17)$     & $(0.15)$     \\
\hline
AIC                             & $18185.44$    & $17853.24$    & $34087.11$   & $33705.72$   \\
Log Likelihood                  & $-9084.72$    & $-8918.62$    & $-17035.55$  & $-16844.86$  \\
Num. obs.                       & $1907$        & $1907$        & $1895$       & $1895$       \\
Num. groups: Receiving country     & $156$         &             & $156$        &            \\
Var: Receiving country (Intercept) & $0.69$        &             & $0.65$       &            \\
Num. groups: Sending country       &             & $163$         &            & $162$        \\
Var: Sending country (Intercept)   &            & $1.35$        &            & $1.49$       \\
\hline
\multicolumn{5}{l}{\scriptsize{$^{***}p<0.001$; $^{**}p<0.01$; $^{*}p<0.05$}}
\end{tabular}

\end{center}
\end{table}

\begin{table}[H]
\caption{Negative Binomial Multilevel models: Gender Inequality Index in sending countries and visibility. Estimating difference in visibility measured as retweet and follower count. Ambassadors are nested in their respective receiving countries (Models 1 and 3) and sending countries (Models 2 and 4). The coefficients reflect log change in the dependent variables per unit change in the independent variables.}
\begin{center}
\begin{tabular}{l c c c c}
\tabularnewline\midrule\midrule
Dependent Variables:&\multicolumn{2}{c}{Retweets}&\multicolumn{2}{c}{Followers}\\
Model:&(1) & (2) & (3) & (4)\\
\hline
Intercept                       & $-0.67^{**}$  & $-0.80^{**}$  & $6.15^{***}$ & $5.77^{***}$ \\
                                & $(0.21)$      & $(0.25)$      & $(0.16)$     & $(0.24)$     \\
in-degree (receiving country)    & $1.50^{***}$  & $1.19^{***}$  & $1.54^{***}$ & $1.71^{***}$ \\
                                & $(0.41)$      & $(0.18)$      & $(0.36)$     & $(0.15)$     \\
Woman                           & $-0.85^{***}$ & $-0.72^{***}$ & $-0.23^{*}$  & $-0.32^{**}$ \\
                                & $(0.13)$      & $(0.12)$      & $(0.11)$     & $(0.10)$     \\
Gender Inequality Index         & $-0.30^{**}$  & $-0.63^{*}$   & $0.23^{*}$   & $-0.09$      \\
                                & $(0.11)$      & $(0.26)$      & $(0.09)$     & $(0.27)$     \\
log(n tweets + 0.1)             & $1.16^{***}$  & $1.16^{***}$  & $0.36^{***}$ & $0.31^{***}$ \\
                                & $(0.02)$      & $(0.03)$      & $(0.01)$     & $(0.01)$     \\
Woman x Gender Inequality Index & $0.60^{**}$   & $0.64^{***}$  & $0.14$       & $0.31$       \\
                                & $(0.20)$      & $(0.19)$      & $(0.17)$     & $(0.16)$     \\
\hline
AIC                             & $18282.18$    & $18020.81$    & $34164.88$   & $33809.45$   \\
Log Likelihood                  & $-9133.09$    & $-9002.41$    & $-17074.44$  & $-16896.72$  \\
Num. obs.                       & $1912$        & $1912$        & $1900$       & $1900$       \\
Num. groups: Receiving country     & $172$         &             & $172$        &            \\
Var: Receiving country (Intercept) & $0.87$        &             & $0.67$       &            \\
Num. groups: Sending country       &             & $145$         &           & $144$        \\
Var: Sending country (Intercept)   &             & $1.29$        &            & $1.50$       \\
\hline
\multicolumn{5}{l}{\scriptsize{$^{***}p<0.001$; $^{**}p<0.01$; $^{*}p<0.05$}}
\end{tabular}

\label{table:nbm-gii}
\end{center}
\end{table}

\begin{table}[H]
\caption{Negative Binomial models: Visibility (retweet and follower count) and \% of women in parliament in receiving country. Ambassadors are nested in their respective receiving countries (Models 1 and 3) and sending countries (Models 2 and 4). The coefficients reflect log change in the dependent variables per unit change in the independent variables.}
\begin{center}
\begin{tabular}{l c c c c}
\tabularnewline\midrule\midrule
Dependent Variables:&\multicolumn{2}{c}{Retweets}&\multicolumn{2}{c}{Followers}\\
Model:&(1) & (2) & (3) & (4)\\
\hline
Intercept                         & $-0.77^{***}$ & $-0.92^{***}$ & $6.29^{***}$ & $5.81^{***}$ \\
                                  & $(0.22)$      & $(0.19)$      & $(0.17)$     & $(0.15)$     \\
in-degree (receiving country)      & $1.70^{***}$  & $1.19^{***}$  & $1.60^{***}$ & $1.67^{***}$ \\
                                  & $(0.42)$      & $(0.18)$      & $(0.37)$     & $(0.15)$     \\
Woman                             & $-0.73^{***}$ & $-0.59^{***}$ & $-0.35^{**}$ & $-0.23^{*}$  \\
                                  & $(0.14)$      & $(0.13)$      & $(0.12)$     & $(0.11)$     \\
\% of women in parliament         & $-0.28$       & $-0.51^{***}$ & $-0.07$      & $-0.16$      \\
                                  & $(0.19)$      & $(0.10)$      & $(0.17)$     & $(0.08)$     \\
log(n tweets + 0.1)               & $1.17^{***}$  & $1.15^{***}$  & $0.36^{***}$ & $0.31^{***}$ \\
                                  & $(0.02)$      & $(0.02)$      & $(0.01)$     & $(0.01)$     \\
Woman x \% of women in parliament & $0.24$        & $0.32$        & $0.29$       & $0.08$       \\
                                  & $(0.20)$      & $(0.17)$      & $(0.17)$     & $(0.15)$     \\
\hline
AIC                               & $18564.57$    & $18235.23$    & $34858.63$   & $34481.58$   \\
Log Likelihood                    & $-9274.29$    & $-9109.62$    & $-17421.32$  & $-17232.79$  \\
Num. obs.                         & $1954$        & $1954$        & $1942$       & $1942$       \\
Num. groups: Receiving country       & $171$         &             & $171$        &            \\
Var: Receiving country (Intercept)   & $0.88$        &             & $0.70$       &            \\
Num. groups: Sending country         &             & $164$         &            & $163$        \\
Var: Sending country (Intercept)     &             & $1.38$        &            & $1.49$       \\
\hline
\multicolumn{5}{l}{\scriptsize{$^{***}p<0.001$; $^{**}p<0.01$; $^{*}p<0.05$}}
\end{tabular}
\label{table:nbm-vis}
\end{center}
\end{table}

\begin{table}[H]
\caption{Negative Binomial Multilevel Models: Visibility (retweet and follower count) and \% of women in parliament in sending country. Ambassadors are nested in their respective receiving countries (Models 1 and 3) and sending countries (Models 2 and 4). The coefficients reflect log change in the dependent variables per unit change in the independent variables.}
\begin{center}
\begin{tabular}{l c c c c}
\tabularnewline\midrule\midrule
Dependent Variables:&\multicolumn{2}{c}{Retweets}&\multicolumn{2}{c}{Followers}\\
Model:&(1) & (2) & (3) & (4)\\
\hline
Intercept                       & $-0.83^{***}$ & $-1.30^{***}$ & $6.35^{***}$ & $5.70^{***}$ \\
                                & $(0.21)$      & $(0.20)$      & $(0.16)$     & $(0.17)$     \\
in-degree (receiving country)    & $1.58^{***}$  & $1.18^{***}$  & $1.55^{***}$ & $1.66^{***}$ \\
                                & $(0.42)$      & $(0.18)$      & $(0.37)$     & $(0.15)$     \\
Woman                           & $-0.43^{**}$  & $-0.07$       & $-0.06$      & $-0.14$      \\
                                & $(0.15)$      & $(0.14)$      & $(0.12)$     & $(0.12)$     \\
Gender Inequality Index         & $-0.10$       & $0.16$        & $-0.14$      & $0.06$       \\
                                & $(0.11)$      & $(0.26)$      & $(0.09)$     & $(0.25)$     \\
log(n tweets + 0.1)             & $1.18^{***}$  & $1.16^{***}$  & $0.36^{***}$ & $0.31^{***}$ \\
                                & $(0.02)$      & $(0.02)$      & $(0.01)$     & $(0.01)$     \\
Woman x Gender Inequality Index & $-0.31$       & $-0.67^{***}$ & $-0.21$      & $-0.09$      \\
                                & $(0.20)$      & $(0.18)$      & $(0.17)$     & $(0.16)$     \\
\hline
AIC                             & $18626.52$    & $18318.29$    & $34955.55$   & $34586.02$   \\
Log Likelihood                  & $-9305.26$    & $-9151.15$    & $-17469.78$  & $-17285.01$  \\
Num. obs.                       & $1959$        & $1959$        & $1947$       & $1947$       \\
Num. groups: Receiving country     & $172$         & $$            & $172$        & $$           \\
Var: Receiving country (Intercept) & $0.89$        & $$            & $0.69$       & $$           \\
Num. groups: Sending country       & $$            & $163$         & $$           & $162$        \\
Var: Sending country (Intercept)   & $$            & $1.43$        & $$           & $1.50$       \\
\hline
\multicolumn{5}{l}{\scriptsize{$^{***}p<0.001$; $^{**}p<0.01$; $^{*}p<0.05$}}
\end{tabular}
\label{table:nbb-vis-sending}
\end{center}
\end{table}

\begin{table}[H]
\caption{Negative Binomial Multilevel Models: Visibility (retweet and follower count) and Gender Social Norms Index in the receiving country. Ambassadors are nested in their respective receiving countries (Models 1 and 3) and sending countries (Model 2 and 4). The coefficients reflect log change in the dependent variables per unit change in the independent variables.}
\begin{center}
\begin{tabular}{l c c c c}
\tabularnewline\midrule\midrule
Dependent Variables:&\multicolumn{2}{c}{Retweets}&\multicolumn{2}{c}{Followers}\\
Model:&(1) & (2) & (3) & (4)\\
\hline
Intercept                         & $-1.33^{***}$ & $-2.29^{***}$ & $5.93^{***}$ & $5.36^{***}$ \\
                                  & $(0.38)$      & $(0.25)$      & $(0.33)$     & $(0.20)$     \\
in-degree (receiving country)      & $2.49^{***}$  & $2.07^{***}$  & $1.85^{***}$ & $2.04^{***}$ \\
                                  & $(0.59)$      & $(0.25)$      & $(0.54)$     & $(0.21)$     \\
Woman                             & $-0.78^{***}$ & $-0.46^{**}$  & $-0.01$      & $-0.10$      \\
                                  & $(0.18)$      & $(0.16)$      & $(0.14)$     & $(0.14)$     \\
Gender Social Norms Index         & $0.33$        & $0.57^{***}$  & $0.34$       & $0.15$       \\
                                  & $(0.26)$      & $(0.13)$      & $(0.24)$     & $(0.11)$     \\
log(n tweets + 0.1)               & $1.15^{***}$  & $1.24^{***}$  & $0.35^{***}$ & $0.31^{***}$ \\
                                  & $(0.03)$      & $(0.03)$      & $(0.02)$     & $(0.02)$     \\
Woman x Gender Social Norms Index & $0.10$        & $-0.04$       & $-0.17$      & $0.03$       \\
                                  & $(0.27)$      & $(0.22)$      & $(0.21)$     & $(0.19)$     \\
\hline
AIC                               & $11829.49$    & $11495.80$    & $21938.01$   & $21649.26$   \\
Log Likelihood                    & $-5906.74$    & $-5739.90$    & $-10961.01$  & $-10816.63$  \\
Num. obs.                         & $1220$        & $1220$        & $1209$       & $1209$       \\
Num. groups: Receiving country       & $72$          &            & $72$         &            \\
Var: Receiving country (Intercept)   & $0.68$        &             & $0.62$       &            \\
Num. groups: Sending country         &             & $159$         &            & $158$        \\
Var: Sending country (Intercept)     &             & $1.64$        &            & $1.68$       \\
\hline
\multicolumn{5}{l}{\scriptsize{$^{***}p<0.001$; $^{**}p<0.01$; $^{*}p<0.05$}}
\end{tabular}
\label{table:vis-gsn}
\end{center}
\end{table}

\begin{table}[H]
\caption{Negative Binomial Multilevel Models: Visibility (retweet and follower count) and Gender Social Norms Index in the sending country. Ambassadors are nested in their respective receiving countries (Models 1 and 3) and sending countries (Models 2 and 4). The coefficients reflect log change in the dependent variables per unit change in the independent variables.}
\begin{center}
\begin{tabular}{l c c c c}
\tabularnewline\midrule\midrule
Dependent Variables:&\multicolumn{2}{c}{Retweets}&\multicolumn{2}{c}{Followers}\\
Model:&(1) & (2) & (3) & (4)\\
\hline
Intercept                         & $-0.76^{***}$ & $-0.88^{**}$  & $6.58^{***}$  & $6.54^{***}$  \\
                                  & $(0.22)$      & $(0.33)$      & $(0.18)$      & $(0.32)$      \\
in-degree (receiving country)      & $1.21^{**}$   & $1.37^{***}$  & $1.73^{***}$  & $1.68^{***}$  \\
                                  & $(0.42)$      & $(0.20)$      & $(0.40)$      & $(0.19)$      \\
Woman                             & $-0.70^{***}$ & $-0.56^{***}$ & $-0.56^{***}$ & $-0.48^{***}$ \\
                                  & $(0.14)$      & $(0.13)$      & $(0.13)$      & $(0.12)$      \\
Gender Social Norms Index         & $0.27^{*}$    & $-0.22$       & $-0.36^{**}$  & $-0.52$       \\
                                  & $(0.13)$      & $(0.35)$      & $(0.11)$      & $(0.36)$      \\
log(n tweets + 0.1)               & $1.16^{***}$  & $1.15^{***}$  & $0.35^{***}$  & $0.30^{***}$  \\
                                  & $(0.03)$      & $(0.03)$      & $(0.02)$      & $(0.02)$      \\
Woman x Gender Social Norms Index & $-0.02$       & $0.21$        & $0.25$        & $0.23$        \\
                                  & $(0.24)$      & $(0.22)$      & $(0.21)$      & $(0.20)$      \\
\hline
AIC                               & $12899.19$    & $12701.10$    & $23527.82$    & $23409.92$    \\
Log Likelihood                    & $-6441.60$    & $-6342.55$    & $-11755.91$   & $-11696.96$   \\
Num. obs.                         & $1289$        & $1289$        & $1280$        & $1280$        \\
Num. groups: Receiving country       & $169$         &             & $169$         &             \\
Var: Receiving country (Intercept)   & $0.79$        &             & $0.79$        &             \\
Num. groups: Sending country         &             & $71$          &            & $71$          \\
Var: Sending country (Intercept)     &             & $1.17$        &             & $1.26$        \\
\hline
\multicolumn{5}{l}{\scriptsize{$^{***}p<0.001$; $^{**}p<0.01$; $^{*}p<0.05$}}
\end{tabular}

\label{table:nbb-vis-gsni}
\end{center}
\end{table}

\begin{table}[H]
\caption{Negative Binomial Multilevel models: difference in visibility (retweet and follower count) and prestige (receiving country in-degree). Ambassadors are nested in their respective receiving countries (Models 1 and 3) and sending countries (Models 2 and 4). The coefficients reflect log change in the dependent variables per unit change in the independent variables.}
\begin{center}
\begin{tabular}{l c c c c}
\tabularnewline\midrule\midrule
Dependent Variables:&\multicolumn{2}{c}{Retweets}&\multicolumn{2}{c}{Followers}\\
Model:&(1) & (2) & (3) & (4)\\
\midrule \emph{Variables}&   &   &   &  \\
Intercept                       & $-0.65^{***}$ & $-0.95^{***}$ & $6.62^{***}$ & $6.35^{***}$ \\
                                & $(0.16)$      & $(0.17)$      & $(0.11)$     & $(0.13)$     \\
Woman                           & $-0.34^{*}$   & $-0.24^{*}$   & $-0.17$      & $-0.26^{*}$  \\
                                & $(0.14)$      & $(0.12)$      & $(0.12)$     & $(0.10)$     \\
Above median in-degree           & $0.81^{***}$  & $0.54^{***}$  & $0.59^{**}$  & $0.41^{***}$ \\
                                & $(0.22)$      & $(0.10)$      & $(0.19)$     & $(0.09)$     \\
log(n tweets + 0.1)             & $1.18^{***}$  & $1.18^{***}$  & $0.36^{***}$ & $0.33^{***}$ \\
                                & $(0.02)$      & $(0.03)$      & $(0.01)$     & $(0.01)$     \\
Woman x Above median in-degree   & $-0.56^{**}$  & $-0.45^{*}$   & $-0.01$      & $0.17$       \\
                                & $(0.20)$      & $(0.18)$      & $(0.17)$     & $(0.16)$     \\
\hline
AIC                             & $18635.79$    & $18350.87$    & $34987.08$   & $34685.55$   \\
Log Likelihood                  & $-9310.90$    & $-9168.43$    & $-17486.54$  & $-17335.77$  \\
Num. obs.                       & $1960$        & $1960$        & $1948$       & $1948$       \\
Num. groups: Receiving country     & $172$         &             & $172$        &            \\
Var: Receiving country (Intercept) & $0.94$        &             & $0.74$       &            \\
Num. groups: Sending country       &             & $164$         &            & $163$        \\
Var: Sending country (Intercept)   &             & $1.39$        &            & $1.49$       \\
\hline
\multicolumn{5}{l}{\scriptsize{$^{***}p<0.001$; $^{**}p<0.01$; $^{*}p<0.05$}}
\end{tabular}
\label{table:reg_vis_mediated_in-degree}
\end{center}
\end{table}

\begin{table}[H] \centering 
  \caption{Simple OLS: testing whether negativity is mediated by Gender Inequality Index (GII), \% of women in parliament and in-degree of the receiving country } 
   \resizebox{\linewidth}{!}{%
  \label{} 
\begin{tabular}{@{\extracolsep{5pt}}lcccccc} 
\\[-1.8ex]\hline 
\hline \\[-1.8ex] 
 & \multicolumn{6}{c}{\textit{Dependent variable:}} \\ 
\cline{2-7} 
\\[-1.8ex] & \multicolumn{3}{c}{\% of negative replies} & \multicolumn{3}{c}{\% of positive replies} \\ 
\\[-1.8ex] & (1) & (2) & (3) & (4) & (5) & (6)\\ 
\hline \\[-1.8ex] 
 Woman & $-$0.035$^{**}$ & $-$0.035$^{**}$ & $-$0.051$^{***}$ & 0.062$^{***}$ & 0.043$^{*}$ & 0.075$^{***}$ \\ 
  & (0.016) & (0.017) & (0.016) & (0.022) & (0.024) & (0.022) \\ 
  & & & & & & \\ 
 Above Median GII & 0.017 &  &  & $-$0.013 &  &  \\ 
  & (0.013) &  &  & (0.018) &  &  \\ 
  & & & & & & \\ 
 Woman x Above Median GII & $-$0.001 &  &  & $-$0.003 &  &  \\ 
  & (0.024) &  &  & (0.033) &  &  \\ 
  & & & & & & \\ 
 Above median \% of women in parliament &  & $-$0.006 &  &  & 0.002 &  \\ 
  &  & (0.013) &  &  & (0.018) &  \\ 
  & & & & & & \\ 
 Woman x Above median \% \\ of women in parliament &  & $-$0.004 &  &  & 0.034 &  \\ 
  &  & (0.024) &  &  & (0.033) &  \\ 
  & & & & & & \\ 
 Above median in-degree  &  &  & 0.008 &  &  & $-$0.004 \\ 
  &  &  & (0.013) &  &  & (0.018) \\ 
  & & & & & & \\ 
 Woman x Above median in-degree &  &  & 0.031 &  &  & $-$0.031 \\ 
  &  &  & (0.024) &  &  & (0.033) \\ 
  & & & & & & \\ 
 Constant & 0.200$^{***}$ & 0.211$^{***}$ & 0.205$^{***}$ & 0.443$^{***}$ & 0.435$^{***}$ & 0.438$^{***}$ \\ 
  & (0.009) & (0.009) & (0.009) & (0.012) & (0.012) & (0.013) \\ 
  & & & & & & \\ 
\hline \\[-1.8ex] 
Observations & 1,424 & 1,424 & 1,424 & 1,424 & 1,424 & 1,424 \\ 
Residual Std. Error (df = 1420) & 0.203 & 0.203 & 0.203 & 0.280 & 0.280 & 0.280 \\ 
F Statistic (df = 3; 1420) & 4.011$^{***}$ & 3.375$^{**}$ & 4.618$^{***}$ & 4.932$^{***}$ & 5.179$^{***}$ & 5.154$^{***}$ \\ 
\hline 
\hline \\[-1.8ex] 
\textit{Note:}  & \multicolumn{6}{r}{$^{*}$p$<$0.1; $^{**}$p$<$0.05; $^{***}$p$<$0.01} \\ 
\end{tabular}}  
\end{table}

\newpage
\addcontentsline{toc}{subsection}{Appendix E: Annotation guide}
\subsection*{Appendix E: Annotation Guide}
\label{sec:app-annotation}

\renewcommand{\thefigure}{E\arabic{figure}}
\setcounter{figure}{0}
\renewcommand{\thetable}{E\arabic{table}}
\setcounter{table}{0}

You are asked to look up diplomatic accounts on Twitter in order to retrieve their Twitter handle name and to classify their publicly displayed gender. Please read the instructions carefully before proceeding.
\newline
The annotation process should be done in three steps:

\begin{enumerate}
\item Search for the diplomat's name on Twitter
\item Retrieve the account handle name.
\item  Annotate their perceived gender.
\end{enumerate}

\subsubsection*{Step 1}

Use your browser to open the link in the ``twitter\_link'' column in the spreadsheet. This will lead you to a window on Twitter where you can search for users with a matching name.
\newline

\subsubsection*{Step 2}
Find the individual in the Twitter search window, open their account, and copy-paste their Twitter handle name (e.g., @USAmbDenmark) into the ``handle'' column. You can only choose one handle name per individual. Please leave the row blank if you cannot find the person or if the account is set to private mode. You are asked to include only the official accounts that appear authentic and belong to the respective diplomatic employees (e.g., ambassadors, and foreign ministers). You are allowed to use Google Translate when in doubt about the content of the profile. 

\subsubsection*{Step 3}
Write down the number that best represents that account's publicly displayed gender based on the gender cues in the Twitter profile. 
\newline
\begin{table}[H]
\begin{tabular}{|l|l|}
\hline
Category & Number \\ \hline
Men     & 0      \\ \hline
Women   & 1      \\ \hline
Other    & 2      \\ \hline
Unclear  & 3      \\ \hline
\end{tabular}
\end{table}

When classifying the publicly displayed gender, please examine the name, profile image, and the profile description text in that order. The three elements should always be viewed together in context.

However, the self-description text should be prioritized even if it conflicts with the profile image and user name. For example, if the diplomats have names that are common among men (e.g., John, Jack), use men gender cues in their profile image and simultaneously describe themselves as ``she'', ``her/hers'', or ``mother to three children'' in the profile text, the gender should be labeled as women. Please indicate in the comment section if the diplomats present themselves as transgender.

\underline{``Other''}
If the individuals explicitly describe themselves as not being exclusively men or women, they should be categorized as ``Other''. This includes, for example, individuals who describe themselves as non-binary, gender fluid, or genderqueer.

\underline{``Unclear''}
Select ``Unclear'' if you are not sure which of the above-mentioned categories to choose. Please briefly explain why the gender is unclear in the ``comment'' column.

\subsubsection*{How to determine account's relevance and authenticity?}

\begin{enumerate}
    \item See if there are multiple accounts that portray themselves as the same person.
    
    \item Check whether the text in the profile description or profile image matches their diplomatic position (e.g., ``ambassador''). For example, an account describing herself only as a ``software engineer at Facebook'' with no reference to the diplomatic position should be ignored unless her profile image indicates that she also has the respective diplomatic position (see below). Do not include the accounts, if they no longer have the relevant positions based on the Twitter profile.
    
    \item Examine the profile picture. Accounts with no profile pictures or irrelevant images (e.g., advertisements) should be skipped. An account with no relevant profile description should still be included, if their profile image indicates their diplomatic occupation (e.g., a portrait in front of the respective flag, an image from a diplomatic meeting). You are allowed to google the individual, if you are in doubt and want to see whether the face on Twitter matches the diplomat officially appointed by the respective country.

    \item Examine the most recent tweets. Skip the account if the tweets seem automated, for example, if they appear to be copy-pasted many times, are posted at unusual intervals (exactly every 5 minutes), or have unusual frequency ( hundreds of tweets per day). Please see @voteforkenneth for an example of an account that does not appear fully authentic. 
\end{enumerate}

If you are in doubt whether the account is relevant or authentic, please include the account in the dataset, while mentioning your doubt in the ``comment'' column.

\subsubsection*{Free search}

In the final stage of the annotation, you may be asked to look up the name freely on Twitter instead of using the hyperlink in the dataset. You are allowed to use free search \underline{only} for a dataset specially made for this task -- you will be informed about this before receiving the dataset.

Below you will find a guide for the free search: 
\begin{itemize}
    \item Search for the full name.
    \item Remove all of the middle names.
    \item Remove abbreviations (eg. ``L.'') or any remaining titles (``General'', ``ret. gen.'').
    \item Keep only the last name and add ``Ambassador'' to the search. 
\end{itemize}

You are allowed to iterate through the different steps freely until you find a match or conclude that there are no matches.

\end{document}